\documentclass[twocolumn]{revtex4}

\pdfoutput=1
\usepackage{hyperref}
\hypersetup{
    colorlinks=false,
    pdfborder={0 0 0},
}
\usepackage{bm}

\usepackage{amsmath}
\usepackage{amssymb}
\usepackage{amsthm}
\usepackage{array}
\usepackage{xy}

\usepackage{graphicx}
\usepackage{subfigure}
\usepackage{epstopdf}
\usepackage{grffile}

\begin{document}

\title{Modeling sequence-specific polymers using anisotropic coarse-grained sites allows 
quantitative comparison with experiment}



\author{Thomas K. Haxton\footnote{{\texttt tkhaxton@lbl.gov}}, Ranjan V. Mannige, Ronald N. Zuckermann, and Stephen Whitelam\footnote{{\texttt swhitelam@lbl.gov}}}

\affiliation{Molecular Foundry, Lawrence Berkeley National Laboratory, Berkeley, CA 94720, USA}

\begin{abstract}
Certain sequences of peptoid polymers (synthetic analogs of peptides) assemble into bilayer nanosheets via a nonequilibrium assembly pathway of adsorption, compression, and collapse at an air-water interface.  As with other large-scale dynamic processes in biology and materials science, understanding 
the details of this supramolecular assembly process
requires a modeling approach that captures behavior on a wide range of length and time scales, from those on which individual sidechains fluctuate to those on which assemblies of polymers evolve. 
Here we demonstrate that a new coarse-grained modeling approach is accurate and computationally efficient enough to do so.
Our approach uses only a minimal number of coarse-grained sites, but retains 
independently fluctuating
orientational degrees of freedom for each site. These orientational degrees of freedom allow us to accurately parameterize both bonded and nonbonded interactions, and to generate all-atom configurations with sufficient accuracy to perform atomic scattering calculations and to interface with all-atom simulations.  We have used this approach to reproduce all available experimental X-ray scattering spectra (for stacked nanosheets, and for peptoids adsorbed at air-water interfaces and in solution),
in order to resolve the microscopic, real-space structures responsible for these Fourier-space features. By interfacing with all-atom simulations, we have also laid the foundations for future multiscale simulations of sequence-specific polymers that communicate in both directions across scales.
\end{abstract}



\maketitle


Many of the most important processes in molecular biology, including allostery~\cite{Collier2013, Cui2008}, enzyme catalysis~\cite{McGeagh2011}, molecular recognition~\cite{Tuffery2012}, protein homeostasis~\cite{England2008}, and nucleic acid metabolism~\cite{Wang2013}, involve the cooperative motion of large but precisely self-assembled~\cite{Bowman2011} biomolecules.  Engineering synthetic materials with similarly sophisticated functionality will require methods to connect the chemical sequence of large molecules (e.g. biomolecules or sequence-defined synthetic polymers) to their self-assembled form and function.  
These methods must span many orders of magnitude in time and space in order to describe atomically detailed interactions and correlated supramolecular motions, both of which contribute to materialsÕ assembly and function.

\begin{figure}
\includegraphics[width=0.5\textwidth]{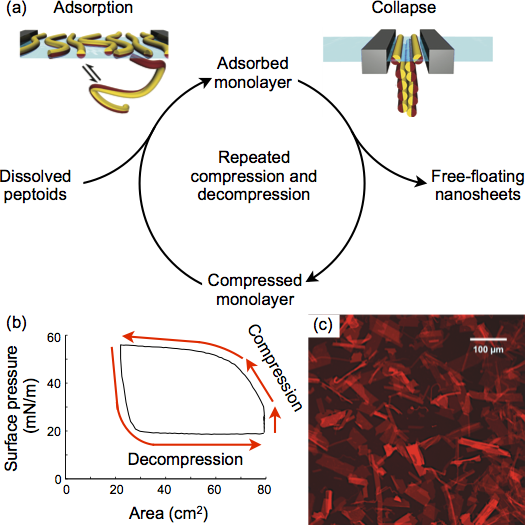}
\caption{(a) Nanosheet production cycle.  Exposed to an air-water interface, amphiphilic peptoid polymers adsorb from solution onto the air-water interface.  Compressing the monolayer induces irreversible collapse into free-floating bilayer nanosheets.  Decompressing the interface completes the cycle, allowing additional peptoids to adsorb.  (b) Experimental plot of the surface pressure and surface area of the monolayer during one production cycle, illustrating the large hysteresis associated with the irreversible formation of nanosheets~\cite{Nam2010, Kudirka2011, Sanii2011, Saniimonolayer}.  (c) Nile red fluorescence micrograph of a solution of free-floating nanosheets~\cite{Olivier2013}.}  
\label{cartoon}
\end{figure}

One class of nanomaterials that show promise as scaffolds for molecular recognition and catalysis~\cite{Olivier2013} are peptoid nanosheets,
solid bilayers that assemble from sequence-defined peptoids (positional isomers of peptides) due to 
a mechanical protocol that acts on many peptoids collectively (Fig.~\ref{cartoon})~\cite{Nam2010, Kudirka2011, Sanii2011, Saniimonolayer}.
Exposed to an air-water interface, amphiphilic peptoid polymers first adsorb from solution onto the interface, forming a structured monolayer. Subsequently compressing the monolayer past a certain pressure induces irreversible collapse into bilayer nanosheets 2.9 nm thick and up to 100 $\mu$m wide.  Such a process is determined by mechanisms operating at multiple length scales: electrostatic interactions at the angstrom scale link sidechains on neighboring polymers; amphiphilic patterning at the 1 nanometer scale allows for adsorption to the air-water interface; the motion of polymers on scales up to their full 10 nm length determines whether in-plane ordering occurs; and nanosheets can extend to scales of order 100 $\mu$m. The associated time scales range from picoseconds, for atomic and molecular fluctuations, to the 
seconds or minutes
on which nanosheets are produced.


Developing a detailed, real-space picture of such a multiscale process requires a modeling approach able to account for mechanisms operating on a broad range of length and time scales.  Studies of macromolecules such as nucleic acids and proteins have shown that coarse-grained modeling can in principle span scales efficiently, by representing explicitly only the most important molecular degrees of freedom, and representing implicitly other degrees of freedom via `effective' interactions~\cite{Nielson2004, Clementi2008, Murtola2009, Tozzini2010, Trylska2010, Kamerlin2011, Hyeon2011, Takada2012, Shinoda2012, Saunders2013, Noid2013}.  
However, reducing the number of degrees of freedom unavoidably discards information,
so the resulting coarse-grained model cannot capture all aspects of the underlying all-atom system.  
For example, it has been shown generally that coarse-grained models parameterized to reproduce all-atom pair distribution functions cannot correctly reproduce thermodynamic properties like energy and pressure, and vice versa~\cite{Louis2002, Stillinger2002, Johnson2007}.



To mitigate this \textit{representability} problem~\cite{Louis2002, Stillinger2002, Johnson2007}, careful choices must be made in 
the two key aspects of a coarse-graining scheme: the choice of which degrees of freedom (or `sites') to be retained, and how interactions between sites should be parameterized.  Most work has focused on the latter, resulting in the development of rigorous interaction-parameterization schemes to target \textit{particular} features (typically distribution functions, forces, or energies) of related all-atom simulations~\cite{Izvekov2005, Noid2008, Noid2008b, Izvekov2010, Muller2011}.

Fewer authors have investigated the choice of which degrees of freedom to retain.  Coarse-grained models based on rigorous parameterization schemes have employed isotropic (spherically symmetric) interactions, for which these schemes are most tractable.  However, it has been shown that the accuracy of isotropic coarse-grained models declines as the underlying all-atom system becomes more anisotropic, due to the fact that more information is lost when averaging spherically over anisotropic interactions than over isotropic ones~\cite{Kowalczyk2009, Kowalczyk2011}.

One way to improve the accuracy of a coarse-grained model with spherically symmetric sites is to increase the number of sites, a strategy often employed in protein modeling~\cite{Monticelli2008, Hills2010, Kar2013}.  For instance, by including between 3 and 8 sites per amino acid residue, the PRIMO protein model~\cite{Kar2013} can estimate all-atom configurations with such accuracy that
all-atom configurations can be passed between PRIMO and atomistic models
with 0.1~\AA~resolution~\cite{Gopal2009}, 
allowing the model to seamlessly interface with all-atom force fields in multiscale simulations~\cite{Predeus2012}. While such an approach clearly represents a protein great detail, the large number of degrees of freedom represented means that simulations are only one order of magnitude faster than all-atom simulations~\cite{Kar2013}.

Here, we 
demonstrate
an alternative strategy for combining accuracy and efficiency within coarse-grained modeling: retain only a minimal number of coarse-grained sites (two per monomer), but include fluctuating orientational degrees of freedom for each site. 
Coarse-grained modelers have recognized the importance of anisotropic interactions in biomolecules, having incorporated directional non-bonded interactions into models for proteins~\cite{Liwo1997, Liwo2004, Yap2008, Majek2009, Alemani2009, Enciso2010, Enciso2012, Spiga2013}, DNA~\cite{Ouldridge2010, Morriss-Andrews2010, Linak2011, Sulc2013}, and lipids~\cite{Orsi2011}.  Efforts in protein modeling have focused on capturing the directionality of backbone hydrogen bonding (absent in peptoids) by approximating the dipole-dipole interaction between peptide groups using only the positions of nearby alpha carbons on the backbone~\cite{Liwo2004, Yap2008, Majek2009, Alemani2009, Enciso2010, Enciso2012}.  In addition, at least one model has accounted for the anisotropic shape of protein side-chains by using ellipsoidal (but energetically isotropic) sidechain sites~\cite{Liwo1997}, and one model has accounted for dipole-dipole interactions between polar sidechains~\cite{Spiga2013}.  Coarse-grained DNA models have focused more on directional nonbonded interactions, including separate base-pairing, base-stacking, and cross-stacking interactions depending on the relative orientation of interacting nucleobases~\cite{Ouldridge2010, Morriss-Andrews2010, Linak2011, Sulc2013}.

Motivated by the success of those strategies, and recognizing the importance of conformational flexibility in peptoids, we created a model with directional interactions depending on \textit{independently fluctuating} orientations, associating each site with both a position and an independent symmetry axis.
As far as we know, the only previous uses of orientational degrees of freedom in coarse-grained biomolecule models are the protein model of Spiga, Alemani, Degiacomi, Cascella, and Dal Peraro, which includes rotating electric dipoles in polar sidechains~\cite{Spiga2013}, and the DNA model of Morriss-Andrews, Rottler, and Plotkin, which includes one floppy orientational degree of freedom per base~\cite{Morriss-Andrews2010}.  
Relative to an isotropic model, 
including a symmetry axis
increases the number of degrees of freedom per site from three to five but improves
 the accuracy of our model in two ways. First, it allows us to parameterize bonded and nonbonded interactions 
that incorporate atomic-level detail.
Second, it allows us to estimate, or \textit{backmap}, all-atom configurations with sufficient accuracy to perform detailed scattering calculations and interface with all-atom simulations.


We plan to use our model to investigate the dynamic, large-scale processes in the nanosheet production cycle (Fig.~\ref{cartoon}): adsorption of solvated peptoid polymers to the air-water interface, ordering of the adsorbed monolayer, and collapse into a free-floating bilayer.  In the current study we establish that our model reproduces all 
known structural features
of the equilibrium states involved in this production cycle,
features measured by
solution X-ray scattering of solvated polymers, grazing-incidence X-ray scattering 
of monolayers, and oriented X-ray scattering of stacks of bilayer nanosheets.
In so doing we provide a microscopic, real-space interpretation of these Fourier-space features.
In addition, we show how our approach can interface with all-atom simulations with an accuracy comparable to computationally more demanding models that possess more degrees of freedom,
laying the groundwork for efficient multiscale simulations~\cite{Ayton2007, Praprotnik2008, Peter2009, Meier2013} of sequence-specific polymers 
that communicate in both directions across scales.


\section{Model and Methods}

\begin{figure*}
\includegraphics[width=\textwidth]{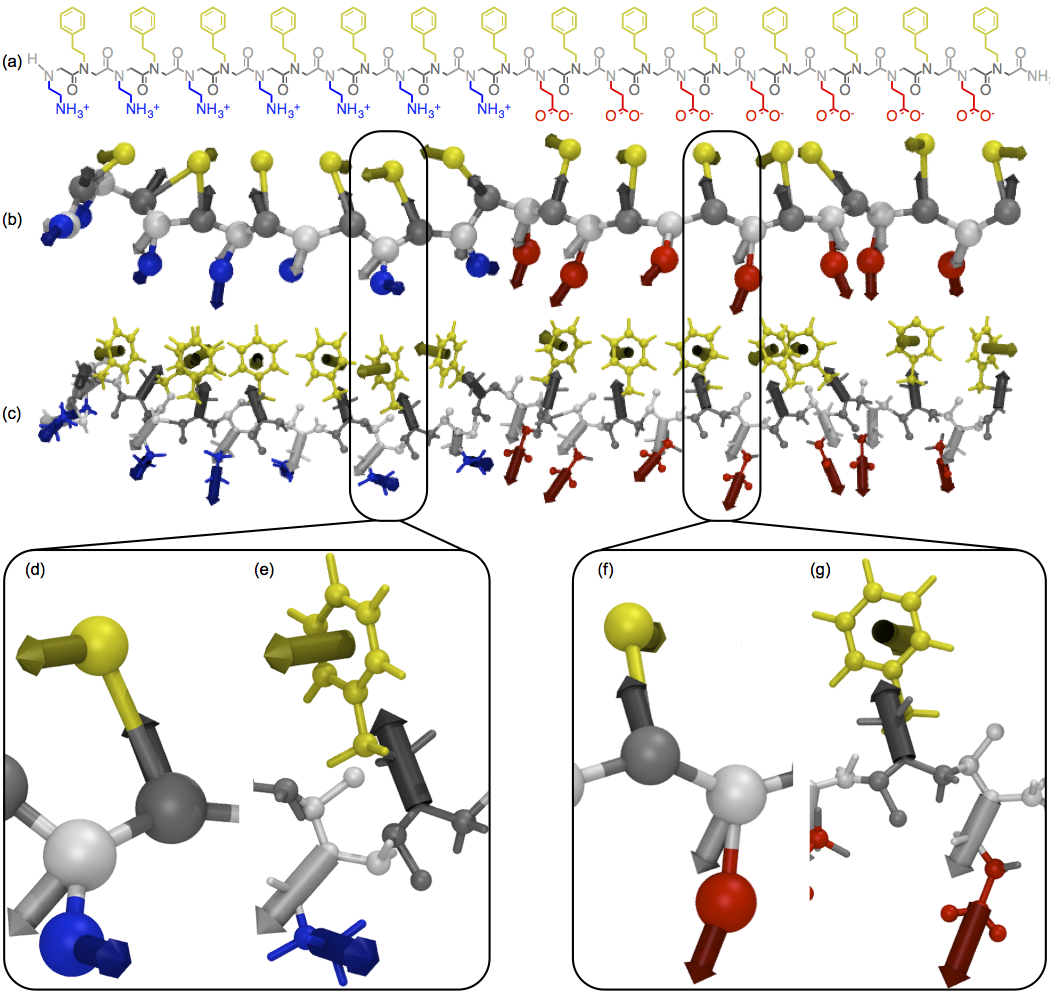}
\caption{(a) Chemical diagram of a 28-monomer ``block-28" peptoid polymer (Nae-Npe)$_{7}$-(Nce-Npe)$_{7}$.  Blue and red parts of the diagram denote positively and negatively charged polar sidechains, yellow parts denote nonpolar sidechains, and grey parts denote the backbone.  (b) Ball-and-stick representation of our coarse-grained peptoid model, taken from a single strand equilibrated within a monolayer. Blue, red, and yellow balls denote positions of the positively charged sidechain sites, grey balls denote positions of the backbone sites, and arrows denote the 
symmetry axis of each site. (c) All-atom representation backmapped from the positions and (superposed) orientations of the coarse-grained sites.  Atoms are colored according to which site they are associated with.  (d-e) Close-up view showing two monomers in the aminoethyl block in (d) coarse-grained and (e) all-atom representations.  (f-g) Close-up view showing two monomers in the carboxyethyl bock in (f) coarse-grained and (g) all-atom representations.}  
\label{schematic}
\end{figure*}

In this paper we focus on the block-charge peptoid illustrated in Fig.~\ref{schematic} (a-c), although we note that our coarse-graining scheme is transferable to different peptoid chemistries and is generalizable to proteins~\cite{Haxtonproteininprep}).  ``Block-$n$" peptoids are (poly)peptoids of the form ((Nae--Npe)$_{n/4}$--(Nce--Npe)$_{n/4}$), built from two equal-length blocks, each consisting of nonpolar $N$-(2-phenylethyl)glycine monomers alternating with charged monomers.  The first block contains positively charged $N$-(2-aminoethyl)glycine monomers, and the second block contains negatively charged $N$-(2-carboxyethyl)glycine monomers.  We investigated ``block-$n$" peptoids of several lengths but present results primarily for block-28, for which the most extensive set of experimental data exists. As illustrated in Fig.~\ref{schematic} (b), we assign two coarse-grained sites per monomer, one for the backbone and one for the sidechain.  The resulting bonded network is a linear chain of $n$ backbone sites, each branched with one sidechain site, plus (optionally) an additional backbone site representing the relatively bulky carboxy-terminus.  We associate the backbone sites with the backbone atoms of each monomer plus the first methylene bridge of the sidechain.  We associate the phenylethyl, aminoethyl, and carboxyethyl sidechain sites with the 
remaining atoms of the sidechains.  
With these associations, the backbone maps to the molecule N-methylacetamide and the sidechain sites map to the ``sidechain analogs"~\cite{Wolfenden1981} toluene, methylammonium, and acetate.  


We characterize each site $i$ by its coordinates $X_i=\{\vec{r}_i, \hat{n}_i\}$, where $\vec{r}_i$ is the site location and $\hat{n}_i$ is the dominant symmetry axis characterizing the atoms associated with each site.
For the backbone sites, mapping $\vec{r}_i$ to the backbone N location and $\hat{n}_i$ to the first sidechain (N-C$_\beta$) bond allows for a straightforward decomposition of bonded interactions into backbone and sidechain terms, as discussed in Section 1 of the Supplemental Methods.  For the phenylethyl sidechain site, mapping $\vec{r}_i$ to the center of the aromatic ring and $\hat{n}_i$ to its normal respects the aromatic ring's axial symmetry (approximating it's 6-fold symmetry as continuous).  For the aminoethyl and carboxyethyl sidechains, mapping $\vec{r}_i$ to the center of mass of the associated atoms and $\hat{n}_i$ to the direction of the central bond (C$_\gamma$-N$_{\rm amino}$ or C$_\gamma$-C$_{\rm carboxy}$) aligns the site with the dipole moment of the associated atoms and leaves only hydrogens and the carboxy oxygens off the symmetry axis.

We write the potential energy of the system as a sum of bonded, solvation, and nonbonded terms,
\begin{equation}
U(\left\{X_i\right\})=U_{\rm b}(\{X_i\})+\sum_{i}U_{\rm s}(r_{iz})+\sum_{i\ne j}U_{\rm nb}(X_i, X_j),
\end{equation}
where the bonded interaction $U_{\rm b}$ includes both two- and three-body terms.
In future work, we plan to allow the shape of the air-water interface to fluctuate, including a fourth potential energy term to couple these fluctuations to the surface tension.  We expect that including these fluctuations will be necessary to allow investigation of the mechanism by which the monolayer collapses into a bilayer. 

We parameterized the potential energy function in two stages.  First, as discussed in Sections 1-3 of the Supplemental Methods, we parameterized the potential energy function to match experimental free energies and all-atom potentials of mean force for relevant small molecules: we parameterized the solvation term to reproduce experimental solvation free energies and all-atom interfacial solvation free energies for the molecular analogs; we parameterized the nonbonded term to reproduce all-atom potentials of mean force for the molecular analogs under the appropriate solvation conditions; and we parameterized the bonded interactions to reproduce all-atom potentials of mean force for representative small molecules.  
Although the use of molecular analog solvation free energies to predict macromolecule free energies has been criticized because it neglects solvent exclusion and ``self-solvation"~\cite{Konig2013}, here we explicitly account for those effects by including both solvation interactions and solvent-mediated nonbonded interactions. 
Using anisotropic coarse-grained sites for these parameterization allowed us to capture atomic-level details in the interactions, such as the torsional flexibility of backbone covalent bonds, without integrating equations of motion for atomic sites.  

Second, as discussed in Section 7 of the Supplemental Methods, we fine-tuned the potential energy function to target both experimental and all-atom simulation~\cite{Mannigeinprep} results for bilayers.  We found that it was sufficient to adjust four terms: a term biasing the aminoethyl sidechain to be in a \textit{trans} configuration, the energy scale of the phenyl-phenyl interaction, and the size scales of the phenyl and backbone sites.  These adjustments simultaneously improved the agreement with experimental bilayer X-ray scattering spectra and improved the agreement with numerous distribution functions in all-atom bilayer simulations.

We call our model the Molecular Foundry Coarse-grained Model for Peptoids (MF-CG-TOID).  Source code in C for initializing, simulating, and analyzing the model are available as Supporting Information.

\begin{figure}
\includegraphics[width=0.5\textwidth]{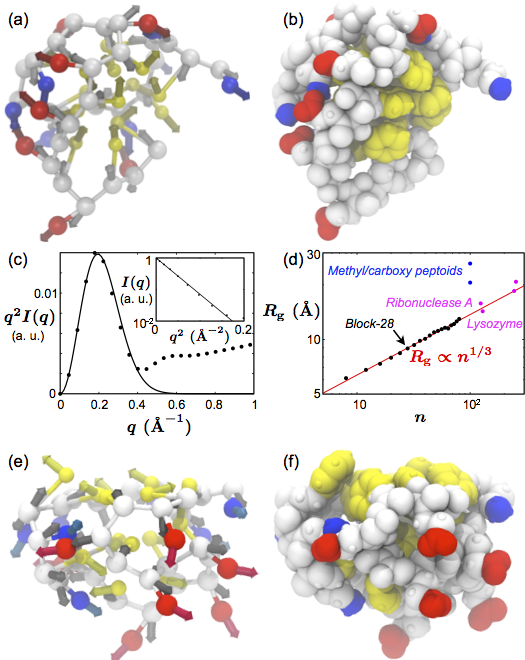}
\caption{(a-b) Snapshots of a single solvated block-28 peptoid in (a) coarse-grained and (b) backmapped all-atom representations.  (c) Kratky (main) and Guinier (inset) plots of the radially averaged X-ray scattering spectrum $I(q)$ for the solvated block-28 peptoid.  The solid curves are the fit to Guinier's law, $I(q)=I_0\exp(-(qR_{\rm g})^2/3)$.  We arbitrarily rescale $I(q)$ by $I_0$.  (d) Radius of gyration $R_{\rm g}$ (from the fit to Guinier's law) vs number of monomers for block-$n$ peptoids.  The red line is a fit to the scaling expected for spherical globules, restricting the fit to $n\ge 16$.  For comparison, the blue points
are experimental values for 100-monomer peptoids with either alternating (top point) or ``protein-like" (bottom point) patterns of 80\% methyl and 20\% carboxyethyl sidechains~\cite{Murnen2012}, and the magenta points are experimental values for globular proteins from Ref.~\cite{Mylonas2007}.  (e-f) Snapshots of the block-28 peptoid exposed to a horizontal air-water interface.  The atoms in (b) and (f) are shown with twice their van der Waals radii to show the exposed aromatic surfaces (yellow).}  
\label{solvated}
\end{figure}

\begin{figure*}
\includegraphics[width=\textwidth]{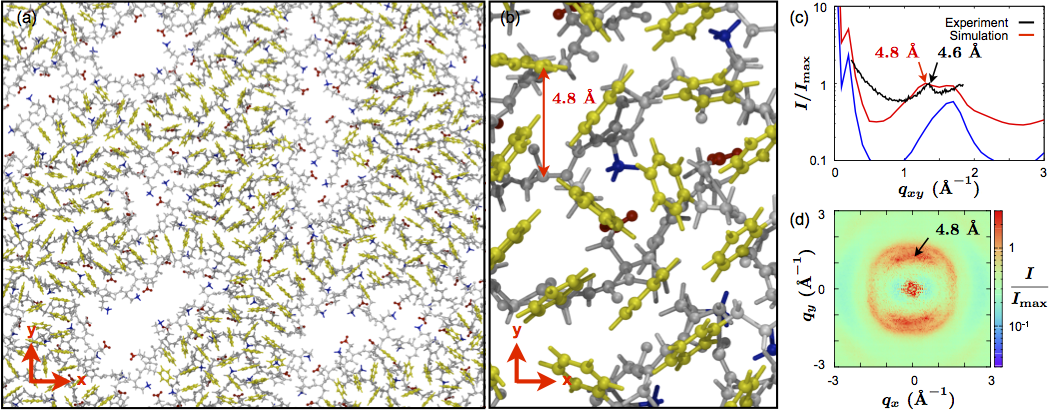}
\caption{(a) Wide-angle and (b) close-up snapshots of a periodic monolayer of 48 block-28 peptoids simulated at the experimental equilibrium surface pressures of 31 mN/m.  The peptoids are stable at the air-water interface, filling most of the interface with small voids near some peptoid termini.  Peptoids remain predominantly parallel to their neighbors, separated by a characteristic distance of 4.8~\AA~(red arrow in (b)).  
(c) Comparison of the experimental grazing-incidence X-ray scattering~\cite{Saniimonolayer} (black) and simulated (red) radially averaged in-plane X-ray spectra show a similar peak location and amplitude at 4.6 (experiment) and 4.8 (simulation) \AA.  The spectra are plotted on a log scale and normalized by the amplitude $I_{\bf max}$ of the dominant peak, in order to allow quantitative comparison without knowing the incident X-ray intensity in experiments (varying which would only shift the log plot up or down).  (d) Two-dimensional in-plane spectrum of the radially anisotropic simulated monolayer confirms that the dominant peak comes from correlations in the $y$ direction.}  
\label{monolayer}
\end{figure*}

We expect that combining our current model with careful treatment of fluctuations of the air-water interface will allow us to investigate the large-scale dynamic processes of adsorption and collapse.  To gain confidence that these studies will connect directly to experiment, we first establish in the current paper that our model captures 
the structural 
properties of the three metastable states relevant for the nanosheet production cycle (Fig.~\ref{cartoon}): solvated polymers, adsorbed monolayers, and free-floating bilayers.  To calculate these properties, we performed Monte Carlo simulations with periodic boundary conditions in the appropriate ensembles, as detailed in Section 4 and 5 of the Supplemental Methods.  
As discussed in Section 6 of the Supplemental Methods, we compared simulated and experimental structures by calculating scattering spectra on all-atom configurations generated from our coarse-grained simulations.
Although all-atom scattering spectra have been generated from other coarse-grained models~\cite{Tschop1998, Shih2007, Perlmutter2009}, previous approaches have included energy minimization and annealing steps to remove unphysical local configurations.  We believe that our work, using accurate all-atom configurations generated by anisotropic coarse-grained sites, is the first example of a coarse-grained model able to generate accurate all-atom scattering spectra directly.

\section{Comparison with Experiment}

\begin{figure*}
\includegraphics[width=\textwidth]{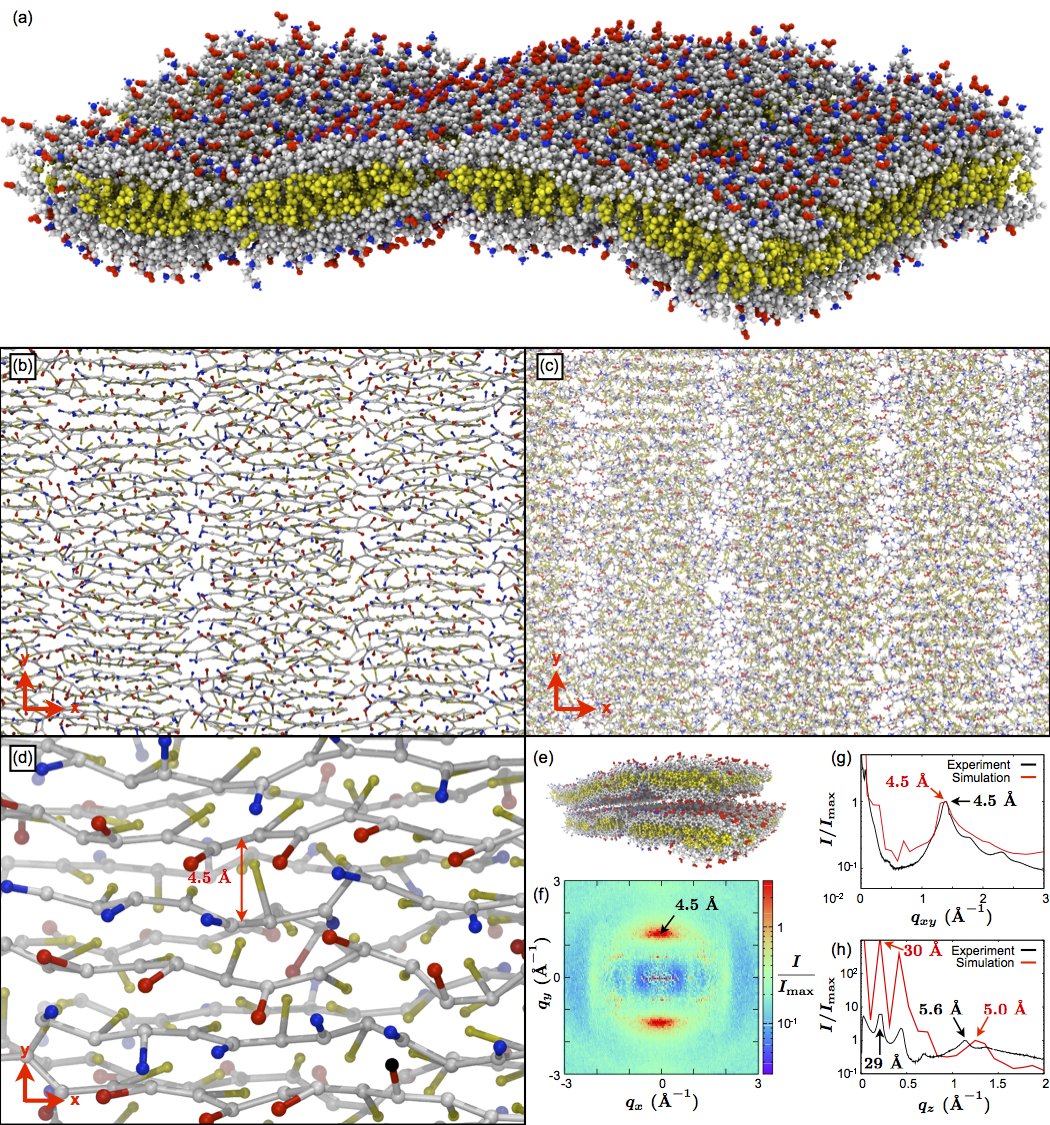}
\caption{(a) All-atom representation of one periodic cell of a bilayer nanosheet at zero tension (96 block-28 peptoids).  (b-c) Wide-angle snapshots looking down on the nanosheet, shown in (a) coarse-grained and (b) all-atom representations.  The bilayer is stable in a rectilinear configuration, with peptoids running in the $x$ direction and seams of terminii running in the $y$ direction.  (d) Close-up snapshots in the coarse-grained representation illustrate the typical 4.5~\AA~spacing between parallel peptoid backbones in each leaf of the bilayer.  (e) We used a periodic cell of two stacked bilayers to calculate the in-plane and out-of-plane X-ray spectra, in order to compare to those obtained experimentally~\cite{Saniimonolayer} from stacked bilayers.  (f) Two-dimensional simulated in-plane X-ray spectrum reveals a dominant peak in the $y$ direction, corresponding to a typical spacing of 4.5~\AA~between parallel peptoids.  (g) The radially averaged in-plane spectrum agrees well with the experimental spectrum.  (h) Experimental and simulated transverse X-ray spectra exhibit dominant peaks at 29~\AA~and 30~\AA, respectively, corresponding to the lamellar spacing between stacked nanosheets, as well as peaks at 5.6~\AA~and 5.0~\AA, respectively, whose origin is more subtle~\cite{Saniimonolayer}.  The stacking peaks are larger in simulation due the perfect stacking in the periodic $z$ direction.}
\label{bilayer}
\end{figure*}

We start our exploration of the nanosheet production cycle with a study of the equilibrium structure of solvated peptoids. This structure presumably strongly influences the ability of peptoids to adsorb to the air-water interface.  Experimental X-ray~\cite{Murnen2012} and neutron~\cite{Rosales2012, Murnen2013} scattering has indicated that solvated peptoids
tend to collapse into single-chain globules, except for highly charged peptoids 
that form extended conformations
in low ionic strength solutions~\cite{Murnen2013}. 
Our model block peptoid collapses into a single-chain globule
when equilibrated in water because of the strong attraction between aromatic rings in phenylethyl sidechains, as seen in experiments and simulations of other synthetic polymers with suitably designed amphiphilic patterns~\cite{Lozinsky2006, Altintas2012, Akagi2012, Morena2013, LoVerso2014, Terashima2014}.
Snapshots of the block-28 peptoid at the end of the simulation (Fig.~\ref{solvated} (a-b)) show its roughly spherical shape, with most of the yellow aromatic sidechains sequestered in the core.  
The Gaussian peak in the the main panel of Fig.~\ref{solvated} (c) indicates a roughly spherical shape characterized by a radius of gyration $R_{\rm g}$.  The dependence on $n$ for $n$-monomer block peptoids (Fig.~\ref{solvated} (d)) indicates that, for large peptoids, the globule size grows with molecular weight as expected (red line) for spherical globules; peptoids smaller than block-16 do not pack as spherically.  Comparing our simulated results with experimental results for more weakly hydrophobic peptoids consisting of 80\% methyl and 20\% carboxyethyl sidechains~\cite{Murnen2012} (blue points in Fig.~\ref{solvated} (b)) suggests that the negatively charged methyl/carboxy peptoids pack less tightly than the block peptoids, likely due to repulsions between negative charges.  
Instead, the agreement of the red scaling with the magenta data points for globular proteins~\cite{Mylonas2007} suggests that our 50\% phenylethyl peptoids pack as tightly as globular proteins.

When our model peptoid is exposed to an air-water interface, as in the snapshots of Fig.~\ref{solvated} (e-f), part of the hydrophobic core flattens out along the interface as the peptoid strikes a balance between maintaining aromatic interactions within its core, and exposing nonpolar groups to the air. The persistence of the aromatic core may help explain why experimental monolayer formation proceeds slower than would be expected for a diffusion-limited process~\cite{Saniimonolayer}.

The next stage of nanosheet production is the formation at the air-water interface of structured peptoid monolayers, a crucial intermediate between unassociated peptoids and nanosheets~\cite{Sanii2011, Saniimonolayer}. 
Fig.~\ref{monolayer} demonstrates that our coarse-grained model captures the essential 
structural features of the monolayer.  We modeled the bulk monolayer by equilibrating a periodic monolayer at fixed values of the surface pressure~\cite{Saniimonolayer}, including the experimental equilibrium surface pressure $p_{\bf s}=31$ mN/m.
We plan to use our model to explore the equilibrium established between a peptoid solution and an air-water interface; in the current work, we used the surface pressure as a control parameter.  

As shown in Fig.~\ref{monolayer} (a), our simulation box equilibrates as a dense monolayer containing small voids near the terminii of some of the peptoids.  Nonpolar sidechains (yellow) tend to segregate away from these voids.  Inspection of close-up images like Fig.~\ref{monolayer} (b) reveals that peptoids tend to align parallel with their neighbors, separated by a characteristic distance of 4.8~\AA.  The distance 4.8~\AA~appears as a peak in the radially averaged X-ray spectrum (Fig.~\ref{monolayer} (c)).  Both the location and amplitude of the peak agree well with the experimental peak at 4.6~\AA~obtained from grazing incidence X-ray scattering in Ref.~\cite{Saniimonolayer}.  In the simulation, peptoids remain predominantly oriented in the $x$ direction, wrapping around the simulation box, so the simulation sample is radially anisotropic.  This anisotropy allows us to separate the in-plane spectrum into $x$ and $y$ components, helping to confirm that the dominant peak comes from separations (predominantly in the $y$ direction) between parallel peptoids.  The more isotropic halo arises primarily from correlations between nonpolar sidechains, which tend to organize isotropically except for the constraints imposed by their bonds to the backbones.  Calculating the radial average only for atoms within the nonpolar sidechains (blue curve in~\ref{monolayer} (c)) illustrates that this contribution leads to the second peak observed in the simulation, and may explain the peak seen at the right-hand side of the experimental curve (black).


Focusing next on the final part of the nanosheet production cycle, we show in  Fig.~\ref{bilayer} that our model reproduces the key structural features of the bilayer nanosheets themselves. We modeled the interior of a large, free-floating nanosheet by simulating a periodic bilayer at zero tension (Fig.~\ref{bilayer} (a)).  As shown in Fig.~\ref{bilayer} (b-c), the bilayer stabilizes in a rectilinear configuration, with parallel peptoids wrapping around the $x$ direction and seams of terminii running in the $y$ direction.  To calculate X-ray scattering profiles that could be compared with experimental ones, we placed two vertically separated bilayers in a periodic box and allowed them to stack together by allowing the attractive forces between the bilayers to drive out intervening solvent (see Section 6 of the Supplemental Methods).  As shown in Fig.~\ref{bilayer} (e), the bilayers equilibrated at a lamellar spacing of 30~\AA.  Then, we mimicked the experiments of Ref.~\ref{bilayer} by performing X-ray scattering in and out of the bilayer plane.  (Isolated nanosheets showed very little difference in their in-plane spectra.)  

As shown in Fig.~\ref{bilayer} (f), the in-plane spectrum reveals a dominant peak in the $y$ direction corresponding to a separation of 4.5~\AA~between neighboring peptoids.  The distance is somewhat shorter than in the monolayer, probably due to the additional attractive interactions of the bilayer's second leaf, and the peak is more confined to the $y$ direction due to the bilayer's ability to stabilize a rectilinear configuration.  The radially averaged in-plane spectrum agrees very well with experiment (Fig.~\ref{bilayer} (g)).  The agreement is better than for the monolayer because we fine-tuned three model parameters to match this target.  

The transverse spectrum in Fig.~\ref{bilayer} (h) also agrees well with experiment, though the amplitudes of the lamellar peak at 30~\AA~and its two higher harmonics are larger in simulation due to the perfect stacking in the $z$ direction.  As discussed in Ref.~\cite{Saniimonolayer}, the dominant short-range peak at 5.0~\AA~(compared to 5.6~\AA~in the experiments) arises from a subtle combination of many contributions, so the agreement in its shape and location is a good indicator of the quality of the coarse-grained model.


\section{Interfacing with All-Atom Simulations}

The results described above demonstrate the accuracy afforded by our coarse-grained modeling scheme, which can be achieved in a computationally efficient manner. Our coarse-grained model is at least $10^4$-fold faster than an all-atom calculation: the model possesses 10 times fewer interaction sites per unit volume (e.g. 57 coarse-grained sites versus 543 atoms for a block-28 peptoid, not counting the water molecules necessary for the all-atom simulation); it therefore requires 100 times fewer calculations per unit time per unit volume; and it can be propagated using a time step roughly 100 times larger than is used for all-atom calculations ($10^{-13}$ seconds versus $10^{-15}$ seconds; see Section 4 of the Supplemental Methods).

Furthermore, we are able to pass configurations between coarse-grained and all-atom simulations so as to access the efficiency of the coarse-grained representation and the high resolution of the atomic-scale model. Such multiscale simulation has been done by embedding small all-atom regions of interest within coarse-grained simulations~\cite{Neri2005, Shi2006, Machado2011, Mamonov2012, diPasquale2012, Predeus2012}, by performing replica-exchange simulations across scales~\cite{Lyman2006, Lyman2006b, Christen2006, Moritsugu2010}, or by initializing all-atom simulations with coordinates generated from coarse-grained simulations~\cite{Thogersen2008, Stansfeld2011, Perlmutter2011}.  

\begin{table}
\begin{tabular}{| l | l | l | l | l |}
\hline
Model & Monomer & $N_{\rm Sites}$ & $N_{\textrm{Heavy atoms}}$ & RMSD (\AA) \\
\hline
SICHO/CA & Phe & 2 & 11 & 1.149~\cite{Gopal2009} \\
SICHO/CA & Glu & 2 & 9 & 0.984~\cite{Gopal2009} \\
\hline
PRIMO & Phe & 6 & 11 & 0.059~\cite{Gopal2009} \\
PRIMO & Glu & 6 & 9 & 0.098~\cite{Gopal2009} \\
\hline
MF-CG-TOID & Npe & 2 & 12 & 0.291 \\
MF-CG-TOID & Nae & 2 & 7 & 0.300 \\
MF-CG-TOID & Nce & 2 & 9 & 0.469 \\
\hline
\end{tabular}
\caption{Root mean square displacement per heavy atom between initial and regenerated all-atom configurations for coarse-grained peptoid and protein models.}
\label{RMSD}
\end{table}

To establish the feasibility of this approach for the present model, we determined the compatibility of backmapped all-atom configurations of our coarse-grained model with our previously-published CHARMM-based~\cite{CHARMM} all-atom peptoid forcefield, MFTOID~\cite{Mirijanian2013}.  Following the approach used for the PRIMO protein coarse-grained model~\cite{Gopal2009}, we calculated how far atoms move when mapped ``roundtrip" from an all-atom configuration (AA) to our coarse-grained model (CG) and back to an all-atom configuration (AA).  Since the AA$\to$CG mapping reduces the number of degrees of freedom, and the `backmapping' CG$\to$AA is deterministic, atoms must move during the AA$\rightarrow$CG$\rightarrow$AA roundtrip. Because multiscale simulation schemes are based on either the AA$\rightarrow$CG mapping, the CG$\rightarrow$AA backmapping, or both, these schemes work best when the roundtrip distance moved is as small as possible.  

Using an all-atom bilayer nanosheet as a test case~\cite{Mannigeinprep}, we found that the  `roundtrip'  root mean square displacement (RMSD) per heavy atom is 0.291, 0.300, and 0.469~\AA~for Npe, Nae, and Nce monomers, respectively.  
Table~\ref{RMSD} compares these values to those acquired from reconstructing protein test sets using two coarse-grained protein models: SICHO/CA, which has two isotropic sites per monomer and can generate all-atom configurations using the Molecular Modeling Tools for Structural Biology toolset~\cite{Gopal2009, Feig2004}, and PRIMO, an intermediate-scale model designed specifically to interface directly with all-atom simulations~\cite{Gopal2009}.  Note that the excellent RMSD values for the PRIMO model, on the order of 0.1~\AA, are made possible by using nearly as many coarse-grained sites as heavy atoms (e.g. 6 vs 9 for glutamic acid and 6 vs 11 for phenylalanine).  Our RMSD values lie intermediate between the two models, demonstrating that considerable information that can be stored in the orientations of our coarse-grained sites.  

Ongoing work on a related protein model suggests that storing the full orientation of each coarse-grained site (rather than just the principal symmetry axis) may increase the resolution of a coarse-grained model beyond the intermediate-resolution model: mapping backbone heavy atoms from the Protein Data Bank~\cite{PDB} to the protein model and back again yields an RMSD of 0.050~\AA~\cite{Haxtonproteininprep}.  Although using a full orientation would only require increasing the number of degrees of freedom per site from five to six, we chose not to do so for our peptoid model because of the added complexity introduced for parameterizing the effective interactions.

The accuracy of our coarse-grained model, reflected both in its ability to capture atomic-scale interactions and its ability to generate all-atom configurations, relies on the use of anisotropic coarse-grained sites. Although such anisotropy makes the model roughly six times more costly to simulate than an equivalent model comprising only simple isotropic interactions of a similar range (see Section 4 of the Supplemental Material),
it allows us to estimate all-atom configurations with sufficient accuracy to perform accurate scattering calculations and interface directly with all-atom simulations. As discussed above, producing such configurations with all-atom simulations would result in an approximately $10^4$-fold slowdown.  Producing them with a high-resolution coarse-grained model like PRIMO, with three times as many sites as our coarse-grained model, would result in at least a 9-fold slowdown (due to a 3-fold increase in sites and 3-fold increase in force calculations, plus a shorter time step due to stiffer interactions).  

\section{Conclusion}

We have shown that using 
a coarse-grained model with anisotropic coarse-grained sites
permits efficient and accurate simulation of sequence-specific polymers.
Using a minimal number of coarse-grained sites but including an independently
fluctuating
symmetry axis for each, we are able to efficiently sample ensembles of coarse-grained configurations that map to detailed all-atom configurations with reasonable accuracy.  Although we fine-tuned four parameters to improve agreement with the experimental in-plane X-ray spectrum for the bilayer, other emergent features of the model (the transverse bilayer spectrum and the in-plane monolayer spectrum)
matched experiment without fine-tuning. 
We suggest that future efforts to optimize the accuracy of coarse-graining schemes should consider the symmetry of coarse-grained sites as an important variable.  

We plan to use MF-CG-TOID to investigate the large-scale dynamic processes involved in the peptoid nanosheet production cycle, using our coarse-grained model both as a stand-alone tool and as a key component of a multiscale simulation protocol.  We expect that MF-CG-TOID (augmented by parameterization of the necessary sidechains) may also be useful in creating a new class of protein-mimetic materials based on the precision assembly of sequence-defined peptoid polymers, building on applications already developed in 
therapeutics~\cite{Zuckermann2009},  diagnostics~\cite{Gao2010, Yam2011, Reddy2011}, transfection~\cite{Utku2006, Lobo2003}, and 
antibiotics~\cite{Chongsiriwatana2008}.


\section{Acknowledgement}

This project was funded by the Defense Threat Reduction Agency under Contract No. IACRO-B1144571.  Work at the Molecular Foundry and the National Energy Research Scientific Computing Center was supported by the Office of Science, Office of Basic Energy Sciences, of the U.S. Department of Energy under Contract No. DE-AC02-05CH11231.

\section{Supporting information}

(1) Supplemental Methods describing the model parameterization, Monte Carlo simulations, initialization, generation of all-atom configurations and scattering spectra, and model fine-tuning.  (2) Source code in C for initializing, simulating, and analyzing the model.  This material is available free of charge via the Internet at http://pubs.acs.org/.

\bibliography{MF-CG-TOID.bbl}

\begin{thebibliography}{95}
\expandafter\ifx\csname natexlab\endcsname\relax\def\natexlab#1{#1}\fi
\expandafter\ifx\csname bibnamefont\endcsname\relax
  \def\bibnamefont#1{#1}\fi
\expandafter\ifx\csname bibfnamefont\endcsname\relax
  \def\bibfnamefont#1{#1}\fi
\expandafter\ifx\csname citenamefont\endcsname\relax
  \def\citenamefont#1{#1}\fi
\expandafter\ifx\csname url\endcsname\relax
  \def\url#1{\texttt{#1}}\fi
\expandafter\ifx\csname urlprefix\endcsname\relax\def\urlprefix{URL }\fi
\providecommand{\bibinfo}[2]{#2}
\providecommand{\eprint}[2][]{\url{#2}}

\bibitem[{\citenamefont{Collier and Ortiz}(2013)}]{Collier2013}
\bibinfo{author}{\bibfnamefont{G.}~\bibnamefont{Collier}} \bibnamefont{and}
  \bibinfo{author}{\bibfnamefont{V.}~\bibnamefont{Ortiz}},
  \bibinfo{journal}{Arch. Biochem. Biophys.} \textbf{\bibinfo{volume}{538}},
  \bibinfo{pages}{6} (\bibinfo{year}{2013}).

\bibitem[{\citenamefont{Cui and Karplus}(2008)}]{Cui2008}
\bibinfo{author}{\bibfnamefont{Q.}~\bibnamefont{Cui}} \bibnamefont{and}
  \bibinfo{author}{\bibfnamefont{M.}~\bibnamefont{Karplus}},
  \bibinfo{journal}{Protein Sci.} \textbf{\bibinfo{volume}{17}},
  \bibinfo{pages}{1295} (\bibinfo{year}{2008}).

\bibitem[{\citenamefont{McGeagh et~al.}(2011)\citenamefont{McGeagh, Ranaghan,
  and Mulholland}}]{McGeagh2011}
\bibinfo{author}{\bibfnamefont{J.~D.} \bibnamefont{McGeagh}},
  \bibinfo{author}{\bibfnamefont{K.~E.} \bibnamefont{Ranaghan}},
  \bibnamefont{and} \bibinfo{author}{\bibfnamefont{A.~J.}
  \bibnamefont{Mulholland}}, \bibinfo{journal}{Biochim. Biophys. Acta}
  \textbf{\bibinfo{volume}{1814}}, \bibinfo{pages}{1077}
  (\bibinfo{year}{2011}).

\bibitem[{\citenamefont{Tuffery and Derreumaux}(2012)}]{Tuffery2012}
\bibinfo{author}{\bibfnamefont{P.}~\bibnamefont{Tuffery}} \bibnamefont{and}
  \bibinfo{author}{\bibfnamefont{P.}~\bibnamefont{Derreumaux}},
  \bibinfo{journal}{J. R. Soc. Interface} \textbf{\bibinfo{volume}{9}},
  \bibinfo{pages}{20} (\bibinfo{year}{2012}).

\bibitem[{\citenamefont{England et~al.}(2008)\citenamefont{England, Lucent, and
  Pande}}]{England2008}
\bibinfo{author}{\bibfnamefont{J.}~\bibnamefont{England}},
  \bibinfo{author}{\bibfnamefont{D.}~\bibnamefont{Lucent}}, \bibnamefont{and}
  \bibinfo{author}{\bibfnamefont{V.}~\bibnamefont{Pande}},
  \bibinfo{journal}{Curr. Opin. Struct. Biol.} \textbf{\bibinfo{volume}{18}},
  \bibinfo{pages}{163} (\bibinfo{year}{2008}).

\bibitem[{\citenamefont{Wang et~al.}(2013)\citenamefont{Wang, Cui, Hong,
  Paterson, and Laughton}}]{Wang2013}
\bibinfo{author}{\bibfnamefont{H.}~\bibnamefont{Wang}},
  \bibinfo{author}{\bibfnamefont{J.}~\bibnamefont{Cui}},
  \bibinfo{author}{\bibfnamefont{W.}~\bibnamefont{Hong}},
  \bibinfo{author}{\bibfnamefont{I.~C.} \bibnamefont{Paterson}},
  \bibnamefont{and} \bibinfo{author}{\bibfnamefont{C.~A.}
  \bibnamefont{Laughton}}, \bibinfo{journal}{J. Mol. Model}
  \textbf{\bibinfo{volume}{19}}, \bibinfo{pages}{4997} (\bibinfo{year}{2013}).

\bibitem[{\citenamefont{Bowman et~al.}(2011)\citenamefont{Bowman, Voelz, and
  Pande}}]{Bowman2011}
\bibinfo{author}{\bibfnamefont{G.~R.} \bibnamefont{Bowman}},
  \bibinfo{author}{\bibfnamefont{V.~A.} \bibnamefont{Voelz}}, \bibnamefont{and}
  \bibinfo{author}{\bibfnamefont{V.~S.} \bibnamefont{Pande}},
  \bibinfo{journal}{Curr. Opin. Struct. Biol.} \textbf{\bibinfo{volume}{21}},
  \bibinfo{pages}{4} (\bibinfo{year}{2011}).

\bibitem[{\citenamefont{Olivier et~al.}(2013)\citenamefont{Olivier, Cho, Sanii,
  Connolly, Tran, and Zuckermann}}]{Olivier2013}
\bibinfo{author}{\bibfnamefont{G.~K.} \bibnamefont{Olivier}},
  \bibinfo{author}{\bibfnamefont{A.}~\bibnamefont{Cho}},
  \bibinfo{author}{\bibfnamefont{B.}~\bibnamefont{Sanii}},
  \bibinfo{author}{\bibfnamefont{M.~D.} \bibnamefont{Connolly}},
  \bibinfo{author}{\bibfnamefont{H.}~\bibnamefont{Tran}}, \bibnamefont{and}
  \bibinfo{author}{\bibfnamefont{R.}~\bibnamefont{Zuckermann}},
  \bibinfo{journal}{ACS Nano} \textbf{\bibinfo{volume}{7}},
  \bibinfo{pages}{9276} (\bibinfo{year}{2013}).

\bibitem[{\citenamefont{Nam et~al.}(2010)\citenamefont{Nam, Shelby, Choi,
  Marciel, Chen, Tan, Chu, Mesch, Lee, Connolly et~al.}}]{Nam2010}
\bibinfo{author}{\bibfnamefont{K.~T.} \bibnamefont{Nam}},
  \bibinfo{author}{\bibfnamefont{S.~A.} \bibnamefont{Shelby}},
  \bibinfo{author}{\bibfnamefont{P.~H.} \bibnamefont{Choi}},
  \bibinfo{author}{\bibfnamefont{A.~B.} \bibnamefont{Marciel}},
  \bibinfo{author}{\bibfnamefont{R.}~\bibnamefont{Chen}},
  \bibinfo{author}{\bibfnamefont{L.}~\bibnamefont{Tan}},
  \bibinfo{author}{\bibfnamefont{T.~K.} \bibnamefont{Chu}},
  \bibinfo{author}{\bibfnamefont{T.~A.} \bibnamefont{Mesch}},
  \bibinfo{author}{\bibfnamefont{B.-C.} \bibnamefont{Lee}},
  \bibinfo{author}{\bibfnamefont{M.~D.} \bibnamefont{Connolly}},
  \bibnamefont{et~al.}, \bibinfo{journal}{Nature Mater.}
  \textbf{\bibinfo{volume}{9}}, \bibinfo{pages}{454} (\bibinfo{year}{2010}).

\bibitem[{\citenamefont{Kudirka et~al.}(2011)\citenamefont{Kudirka, Tran,
  Sanii, Nam, Choi, Venkateswaran, Chen, Whitelam, and
  Zuckermann}}]{Kudirka2011}
\bibinfo{author}{\bibfnamefont{R.}~\bibnamefont{Kudirka}},
  \bibinfo{author}{\bibfnamefont{H.}~\bibnamefont{Tran}},
  \bibinfo{author}{\bibfnamefont{B.}~\bibnamefont{Sanii}},
  \bibinfo{author}{\bibfnamefont{K.~T.} \bibnamefont{Nam}},
  \bibinfo{author}{\bibfnamefont{P.~H.} \bibnamefont{Choi}},
  \bibinfo{author}{\bibfnamefont{N.}~\bibnamefont{Venkateswaran}},
  \bibinfo{author}{\bibfnamefont{R.}~\bibnamefont{Chen}},
  \bibinfo{author}{\bibfnamefont{S.}~\bibnamefont{Whitelam}}, \bibnamefont{and}
  \bibinfo{author}{\bibfnamefont{R.~N.} \bibnamefont{Zuckermann}},
  \bibinfo{journal}{Peptide Sci.} \textbf{\bibinfo{volume}{96}},
  \bibinfo{pages}{586} (\bibinfo{year}{2011}).

\bibitem[{\citenamefont{Sanii et~al.}(2011)\citenamefont{Sanii, Kudirka, Cho,
  Venkateswaran, Olivier, Olson, Tran, Harada, Tan, and
  Zuckermann}}]{Sanii2011}
\bibinfo{author}{\bibfnamefont{B.}~\bibnamefont{Sanii}},
  \bibinfo{author}{\bibfnamefont{R.}~\bibnamefont{Kudirka}},
  \bibinfo{author}{\bibfnamefont{A.}~\bibnamefont{Cho}},
  \bibinfo{author}{\bibfnamefont{N.}~\bibnamefont{Venkateswaran}},
  \bibinfo{author}{\bibfnamefont{G.~K.} \bibnamefont{Olivier}},
  \bibinfo{author}{\bibfnamefont{A.~M.} \bibnamefont{Olson}},
  \bibinfo{author}{\bibfnamefont{H.}~\bibnamefont{Tran}},
  \bibinfo{author}{\bibfnamefont{R.~M.} \bibnamefont{Harada}},
  \bibinfo{author}{\bibfnamefont{L.}~\bibnamefont{Tan}}, \bibnamefont{and}
  \bibinfo{author}{\bibfnamefont{R.~N.} \bibnamefont{Zuckermann}},
  \bibinfo{journal}{J. Am. Chem. Soc.}  (\bibinfo{year}{2011}).

\bibitem[{\citenamefont{Sanii et~al.}()\citenamefont{Sanii, Haxton, Olivier,
  Barton, Proulx, Whitelam, and Zuckermann}}]{Saniimonolayer}
\bibinfo{author}{\bibfnamefont{B.}~\bibnamefont{Sanii}},
  \bibinfo{author}{\bibfnamefont{T.~K.} \bibnamefont{Haxton}},
  \bibinfo{author}{\bibfnamefont{G.~K.} \bibnamefont{Olivier}},
  \bibinfo{author}{\bibfnamefont{B.}~\bibnamefont{Barton}},
  \bibinfo{author}{\bibfnamefont{C.}~\bibnamefont{Proulx}},
  \bibinfo{author}{\bibfnamefont{S.}~\bibnamefont{Whitelam}}, \bibnamefont{and}
  \bibinfo{author}{\bibfnamefont{R.~N.} \bibnamefont{Zuckermann}},
  \bibinfo{note}{submitted}.

\bibitem[{\citenamefont{Nielson et~al.}(2004)\citenamefont{Nielson, Lopez,
  Srinivas, and Klein}}]{Nielson2004}
\bibinfo{author}{\bibfnamefont{S.~O.} \bibnamefont{Nielson}},
  \bibinfo{author}{\bibfnamefont{C.~F.} \bibnamefont{Lopez}},
  \bibinfo{author}{\bibfnamefont{G.}~\bibnamefont{Srinivas}}, \bibnamefont{and}
  \bibinfo{author}{\bibfnamefont{M.~L.} \bibnamefont{Klein}},
  \bibinfo{journal}{J. Phys. Condens. Matter} \textbf{\bibinfo{volume}{16}},
  \bibinfo{pages}{R481} (\bibinfo{year}{2004}).

\bibitem[{\citenamefont{Clementi}(2008)}]{Clementi2008}
\bibinfo{author}{\bibfnamefont{C.}~\bibnamefont{Clementi}},
  \bibinfo{journal}{Curr. Opin. Struct. Biol.} \textbf{\bibinfo{volume}{18}},
  \bibinfo{pages}{10} (\bibinfo{year}{2008}).

\bibitem[{\citenamefont{Murtola et~al.}(2009)\citenamefont{Murtola, Bunker,
  Vattulainen, Deserno, and Karttunen}}]{Murtola2009}
\bibinfo{author}{\bibfnamefont{T.}~\bibnamefont{Murtola}},
  \bibinfo{author}{\bibfnamefont{A.}~\bibnamefont{Bunker}},
  \bibinfo{author}{\bibfnamefont{I.}~\bibnamefont{Vattulainen}},
  \bibinfo{author}{\bibfnamefont{M.}~\bibnamefont{Deserno}}, \bibnamefont{and}
  \bibinfo{author}{\bibfnamefont{M.}~\bibnamefont{Karttunen}},
  \bibinfo{journal}{Phys. Chem. Chem. Phys.} \textbf{\bibinfo{volume}{11}},
  \bibinfo{pages}{1869} (\bibinfo{year}{2009}).

\bibitem[{\citenamefont{Tozzini}(2010)}]{Tozzini2010}
\bibinfo{author}{\bibfnamefont{V.}~\bibnamefont{Tozzini}}, \bibinfo{journal}{Q.
  Rev. Biophys.} \textbf{\bibinfo{volume}{43}}, \bibinfo{pages}{3}
  (\bibinfo{year}{2010}).

\bibitem[{\citenamefont{Trylska}(2010)}]{Trylska2010}
\bibinfo{author}{\bibfnamefont{J.}~\bibnamefont{Trylska}}, \bibinfo{journal}{J.
  Phys. Condens. Matter} \textbf{\bibinfo{volume}{22}}, \bibinfo{pages}{453101}
  (\bibinfo{year}{2010}).

\bibitem[{\citenamefont{Kamerlin et~al.}(2011)\citenamefont{Kamerlin, Vicatos,
  Dryga, and Warshel}}]{Kamerlin2011}
\bibinfo{author}{\bibfnamefont{S.~C.~L.} \bibnamefont{Kamerlin}},
  \bibinfo{author}{\bibfnamefont{S.}~\bibnamefont{Vicatos}},
  \bibinfo{author}{\bibfnamefont{A.}~\bibnamefont{Dryga}}, \bibnamefont{and}
  \bibinfo{author}{\bibfnamefont{A.}~\bibnamefont{Warshel}},
  \bibinfo{journal}{Annu. Rev. Phys. Chem.} \textbf{\bibinfo{volume}{62}},
  \bibinfo{pages}{41} (\bibinfo{year}{2011}).

\bibitem[{\citenamefont{Hyeon and Thirumalai}(2011)}]{Hyeon2011}
\bibinfo{author}{\bibfnamefont{C.}~\bibnamefont{Hyeon}} \bibnamefont{and}
  \bibinfo{author}{\bibfnamefont{D.}~\bibnamefont{Thirumalai}},
  \bibinfo{journal}{Nat. Commun.} \textbf{\bibinfo{volume}{2}},
  \bibinfo{pages}{487} (\bibinfo{year}{2011}).

\bibitem[{\citenamefont{Takada}(2012)}]{Takada2012}
\bibinfo{author}{\bibfnamefont{S.}~\bibnamefont{Takada}},
  \bibinfo{journal}{Curr. Opin. Struct. Biol.} \textbf{\bibinfo{volume}{22}},
  \bibinfo{pages}{130} (\bibinfo{year}{2012}).

\bibitem[{\citenamefont{Shinoda et~al.}(2012)\citenamefont{Shinoda, DeVane, and
  Klein}}]{Shinoda2012}
\bibinfo{author}{\bibfnamefont{W.}~\bibnamefont{Shinoda}},
  \bibinfo{author}{\bibfnamefont{R.}~\bibnamefont{DeVane}}, \bibnamefont{and}
  \bibinfo{author}{\bibfnamefont{M.~L.} \bibnamefont{Klein}},
  \bibinfo{journal}{Curr. Opin. Struct. Biol.} \textbf{\bibinfo{volume}{22}},
  \bibinfo{pages}{175} (\bibinfo{year}{2012}).

\bibitem[{\citenamefont{Saunders and Voth}(2013)}]{Saunders2013}
\bibinfo{author}{\bibfnamefont{M.~G.} \bibnamefont{Saunders}} \bibnamefont{and}
  \bibinfo{author}{\bibfnamefont{G.~A.} \bibnamefont{Voth}},
  \bibinfo{journal}{Annu. Rev. Biophys.} \textbf{\bibinfo{volume}{42}},
  \bibinfo{pages}{73} (\bibinfo{year}{2013}).

\bibitem[{\citenamefont{Noid}(2013)}]{Noid2013}
\bibinfo{author}{\bibfnamefont{W.~G.} \bibnamefont{Noid}}, \bibinfo{journal}{J.
  Chem. Phys.} \textbf{\bibinfo{volume}{139}}, \bibinfo{pages}{090901}
  (\bibinfo{year}{2013}).

\bibitem[{\citenamefont{Louis}(2002)}]{Louis2002}
\bibinfo{author}{\bibfnamefont{A.~A.} \bibnamefont{Louis}},
  \bibinfo{journal}{J. Phys. Condens. Matter} \textbf{\bibinfo{volume}{14}},
  \bibinfo{pages}{9187} (\bibinfo{year}{2002}).

\bibitem[{\citenamefont{Stillinger et~al.}(2002)\citenamefont{Stillinger,
  Sakai, and Torquato}}]{Stillinger2002}
\bibinfo{author}{\bibfnamefont{F.~H.} \bibnamefont{Stillinger}},
  \bibinfo{author}{\bibfnamefont{H.}~\bibnamefont{Sakai}}, \bibnamefont{and}
  \bibinfo{author}{\bibfnamefont{S.}~\bibnamefont{Torquato}},
  \bibinfo{journal}{J. Chem. Phys.} \textbf{\bibinfo{volume}{117}},
  \bibinfo{pages}{288} (\bibinfo{year}{2002}).

\bibitem[{\citenamefont{Johnson et~al.}(2007)\citenamefont{Johnson,
  Head-Gordon, and Louis}}]{Johnson2007}
\bibinfo{author}{\bibfnamefont{M.~E.} \bibnamefont{Johnson}},
  \bibinfo{author}{\bibfnamefont{T.}~\bibnamefont{Head-Gordon}},
  \bibnamefont{and} \bibinfo{author}{\bibfnamefont{A.~A.} \bibnamefont{Louis}},
  \bibinfo{journal}{J. Chem. Phys.} \textbf{\bibinfo{volume}{126}},
  \bibinfo{pages}{144509} (\bibinfo{year}{2007}).

\bibitem[{\citenamefont{Izvekov and Voth}(2005)}]{Izvekov2005}
\bibinfo{author}{\bibfnamefont{S.}~\bibnamefont{Izvekov}} \bibnamefont{and}
  \bibinfo{author}{\bibfnamefont{G.~A.} \bibnamefont{Voth}},
  \bibinfo{journal}{J. Phys. Chem. B} \textbf{\bibinfo{volume}{109}},
  \bibinfo{pages}{2469} (\bibinfo{year}{2005}).

\bibitem[{\citenamefont{Noid et~al.}(2008{\natexlab{a}})\citenamefont{Noid,
  Chu, Ayton, Krishna, Izkekov, Voth, Das, and Andersen}}]{Noid2008}
\bibinfo{author}{\bibfnamefont{W.~G.} \bibnamefont{Noid}},
  \bibinfo{author}{\bibfnamefont{J.-W.} \bibnamefont{Chu}},
  \bibinfo{author}{\bibfnamefont{G.~S.} \bibnamefont{Ayton}},
  \bibinfo{author}{\bibfnamefont{V.}~\bibnamefont{Krishna}},
  \bibinfo{author}{\bibfnamefont{S.}~\bibnamefont{Izkekov}},
  \bibinfo{author}{\bibfnamefont{G.~A.} \bibnamefont{Voth}},
  \bibinfo{author}{\bibfnamefont{A.}~\bibnamefont{Das}}, \bibnamefont{and}
  \bibinfo{author}{\bibfnamefont{H.~C.} \bibnamefont{Andersen}},
  \bibinfo{journal}{J. Chem. Phys.} \textbf{\bibinfo{volume}{128}},
  \bibinfo{pages}{244114} (\bibinfo{year}{2008}{\natexlab{a}}).

\bibitem[{\citenamefont{Noid et~al.}(2008{\natexlab{b}})\citenamefont{Noid,
  Liu, Wang, Chu, Ayton, Izvekov, Andersen, and Voth}}]{Noid2008b}
\bibinfo{author}{\bibfnamefont{W.~G.} \bibnamefont{Noid}},
  \bibinfo{author}{\bibfnamefont{P.}~\bibnamefont{Liu}},
  \bibinfo{author}{\bibfnamefont{Y.}~\bibnamefont{Wang}},
  \bibinfo{author}{\bibfnamefont{J.-W.} \bibnamefont{Chu}},
  \bibinfo{author}{\bibfnamefont{G.~S.} \bibnamefont{Ayton}},
  \bibinfo{author}{\bibfnamefont{S.}~\bibnamefont{Izvekov}},
  \bibinfo{author}{\bibfnamefont{H.~C.} \bibnamefont{Andersen}},
  \bibnamefont{and} \bibinfo{author}{\bibfnamefont{G.~A.} \bibnamefont{Voth}},
  \bibinfo{journal}{J. Chem. Phys.} \textbf{\bibinfo{volume}{128}},
  \bibinfo{pages}{244115} (\bibinfo{year}{2008}{\natexlab{b}}).

\bibitem[{\citenamefont{Izvekov et~al.}(2010)\citenamefont{Izvekov, Chung, and
  Rice}}]{Izvekov2010}
\bibinfo{author}{\bibfnamefont{S.}~\bibnamefont{Izvekov}},
  \bibinfo{author}{\bibfnamefont{P.~W.} \bibnamefont{Chung}}, \bibnamefont{and}
  \bibinfo{author}{\bibfnamefont{B.~M.} \bibnamefont{Rice}},
  \bibinfo{journal}{J. Chem. Phys.} \textbf{\bibinfo{volume}{133}},
  \bibinfo{pages}{064109} (\bibinfo{year}{2010}).

\bibitem[{\citenamefont{M\"{u}ller}(2011)}]{Muller2011}
\bibinfo{author}{\bibfnamefont{M.}~\bibnamefont{M\"{u}ller}},
  \bibinfo{journal}{J. Stat. Phys.} \textbf{\bibinfo{volume}{145}},
  \bibinfo{pages}{967} (\bibinfo{year}{2011}).

\bibitem[{\citenamefont{Kowalczyk et~al.}(2009)\citenamefont{Kowalczyk, Gauden,
  and Ciach}}]{Kowalczyk2009}
\bibinfo{author}{\bibfnamefont{P.}~\bibnamefont{Kowalczyk}},
  \bibinfo{author}{\bibfnamefont{P.~A.} \bibnamefont{Gauden}},
  \bibnamefont{and} \bibinfo{author}{\bibfnamefont{A.}~\bibnamefont{Ciach}},
  \bibinfo{journal}{J. Phys. Chem. B} \textbf{\bibinfo{volume}{113}},
  \bibinfo{pages}{12988} (\bibinfo{year}{2009}).

\bibitem[{\citenamefont{Kowalczyk et~al.}(2011)\citenamefont{Kowalczyk, Gauden,
  and Ciach}}]{Kowalczyk2011}
\bibinfo{author}{\bibfnamefont{P.}~\bibnamefont{Kowalczyk}},
  \bibinfo{author}{\bibfnamefont{P.~A.} \bibnamefont{Gauden}},
  \bibnamefont{and} \bibinfo{author}{\bibfnamefont{A.}~\bibnamefont{Ciach}},
  \bibinfo{journal}{J. Phys. Chem. B} \textbf{\bibinfo{volume}{115}},
  \bibinfo{pages}{6985} (\bibinfo{year}{2011}).

\bibitem[{\citenamefont{Monticelli et~al.}(2008)\citenamefont{Monticelli,
  Kandasamy, Periole, Larson, Tieleman, and Marrink}}]{Monticelli2008}
\bibinfo{author}{\bibfnamefont{L.}~\bibnamefont{Monticelli}},
  \bibinfo{author}{\bibfnamefont{S.~K.} \bibnamefont{Kandasamy}},
  \bibinfo{author}{\bibfnamefont{X.}~\bibnamefont{Periole}},
  \bibinfo{author}{\bibfnamefont{R.~G.} \bibnamefont{Larson}},
  \bibinfo{author}{\bibfnamefont{D.~P.} \bibnamefont{Tieleman}},
  \bibnamefont{and} \bibinfo{author}{\bibfnamefont{S.-J.}
  \bibnamefont{Marrink}}, \bibinfo{journal}{J. Comput. Chem.}
  \textbf{\bibinfo{volume}{4}}, \bibinfo{pages}{819} (\bibinfo{year}{2008}).

\bibitem[{\citenamefont{Hills et~al.}(2010)\citenamefont{Hills, Lu, and
  Voth}}]{Hills2010}
\bibinfo{author}{\bibfnamefont{R.~D.} \bibnamefont{Hills}},
  \bibinfo{author}{\bibfnamefont{L.}~\bibnamefont{Lu}}, \bibnamefont{and}
  \bibinfo{author}{\bibfnamefont{G.~A.} \bibnamefont{Voth}},
  \bibinfo{journal}{PLOS Comput. Biol.} \textbf{\bibinfo{volume}{6}},
  \bibinfo{pages}{e1000827} (\bibinfo{year}{2010}).

\bibitem[{\citenamefont{Kar et~al.}(2013)\citenamefont{Kar, Gopal, Cheng,
  Predeus, and Feig}}]{Kar2013}
\bibinfo{author}{\bibfnamefont{P.}~\bibnamefont{Kar}},
  \bibinfo{author}{\bibfnamefont{S.~M.} \bibnamefont{Gopal}},
  \bibinfo{author}{\bibfnamefont{Y.-M.} \bibnamefont{Cheng}},
  \bibinfo{author}{\bibfnamefont{A.}~\bibnamefont{Predeus}}, \bibnamefont{and}
  \bibinfo{author}{\bibfnamefont{M.}~\bibnamefont{Feig}}, \bibinfo{journal}{J.
  Chem. Theory Comput.} \textbf{\bibinfo{volume}{9}}, \bibinfo{pages}{3769}
  (\bibinfo{year}{2013}).

\bibitem[{\citenamefont{Gopal et~al.}(2009)\citenamefont{Gopal, Mukherjee,
  Cheng, and Feig}}]{Gopal2009}
\bibinfo{author}{\bibfnamefont{S.~M.} \bibnamefont{Gopal}},
  \bibinfo{author}{\bibfnamefont{S.}~\bibnamefont{Mukherjee}},
  \bibinfo{author}{\bibfnamefont{Y.-M.} \bibnamefont{Cheng}}, \bibnamefont{and}
  \bibinfo{author}{\bibfnamefont{M.}~\bibnamefont{Feig}},
  \bibinfo{journal}{Proteins} \textbf{\bibinfo{volume}{78}},
  \bibinfo{pages}{1266} (\bibinfo{year}{2009}).

\bibitem[{\citenamefont{Predeus et~al.}(2012)\citenamefont{Predeus, Gul, Gopal,
  and Feig}}]{Predeus2012}
\bibinfo{author}{\bibfnamefont{A.~V.} \bibnamefont{Predeus}},
  \bibinfo{author}{\bibfnamefont{S.}~\bibnamefont{Gul}},
  \bibinfo{author}{\bibfnamefont{S.~M.} \bibnamefont{Gopal}}, \bibnamefont{and}
  \bibinfo{author}{\bibfnamefont{M.}~\bibnamefont{Feig}}, \bibinfo{journal}{J.
  Phys. Chem. B} \textbf{\bibinfo{volume}{116}}, \bibinfo{pages}{8610}
  (\bibinfo{year}{2012}).

\bibitem[{\citenamefont{Liwo et~al.}(1997)\citenamefont{Liwo, Oldziej, Pincus,
  Wawak, Rackovsky, and Scheraga}}]{Liwo1997}
\bibinfo{author}{\bibfnamefont{A.}~\bibnamefont{Liwo}},
  \bibinfo{author}{\bibfnamefont{S.}~\bibnamefont{Oldziej}},
  \bibinfo{author}{\bibfnamefont{M.~R.} \bibnamefont{Pincus}},
  \bibinfo{author}{\bibfnamefont{R.~J.} \bibnamefont{Wawak}},
  \bibinfo{author}{\bibfnamefont{S.}~\bibnamefont{Rackovsky}},
  \bibnamefont{and} \bibinfo{author}{\bibfnamefont{H.~A.}
  \bibnamefont{Scheraga}}, \bibinfo{journal}{J. Comput. Chem.}
  \textbf{\bibinfo{volume}{18}}, \bibinfo{pages}{849} (\bibinfo{year}{1997}).

\bibitem[{\citenamefont{Liwo et~al.}(2004)\citenamefont{Liwo, Oldziej,
  Czaplewski, Kozlowska, and Scheraga}}]{Liwo2004}
\bibinfo{author}{\bibfnamefont{A.}~\bibnamefont{Liwo}},
  \bibinfo{author}{\bibfnamefont{S.}~\bibnamefont{Oldziej}},
  \bibinfo{author}{\bibfnamefont{C.}~\bibnamefont{Czaplewski}},
  \bibinfo{author}{\bibfnamefont{U.}~\bibnamefont{Kozlowska}},
  \bibnamefont{and} \bibinfo{author}{\bibfnamefont{H.~A.}
  \bibnamefont{Scheraga}}, \bibinfo{journal}{J. Phys. Chem. B}
  \textbf{\bibinfo{volume}{108}}, \bibinfo{pages}{9421} (\bibinfo{year}{2004}).

\bibitem[{\citenamefont{Yap et~al.}(2008)\citenamefont{Yap, Fawzi, and
  Head-Gordon}}]{Yap2008}
\bibinfo{author}{\bibfnamefont{E.-H.} \bibnamefont{Yap}},
  \bibinfo{author}{\bibfnamefont{N.~L.} \bibnamefont{Fawzi}}, \bibnamefont{and}
  \bibinfo{author}{\bibfnamefont{T.}~\bibnamefont{Head-Gordon}},
  \bibinfo{journal}{Proteins} \textbf{\bibinfo{volume}{70}},
  \bibinfo{pages}{626} (\bibinfo{year}{2008}).

\bibitem[{\citenamefont{M\'ajek and Elber}(2009)}]{Majek2009}
\bibinfo{author}{\bibfnamefont{P.}~\bibnamefont{M\'ajek}} \bibnamefont{and}
  \bibinfo{author}{\bibfnamefont{R.}~\bibnamefont{Elber}},
  \bibinfo{journal}{Proteins} \textbf{\bibinfo{volume}{76}},
  \bibinfo{pages}{822} (\bibinfo{year}{2009}).

\bibitem[{\citenamefont{Alemani et~al.}(2010)\citenamefont{Alemani, Collu,
  Cascella, and Dal~Peraro}}]{Alemani2009}
\bibinfo{author}{\bibfnamefont{D.}~\bibnamefont{Alemani}},
  \bibinfo{author}{\bibfnamefont{F.}~\bibnamefont{Collu}},
  \bibinfo{author}{\bibfnamefont{M.}~\bibnamefont{Cascella}}, \bibnamefont{and}
  \bibinfo{author}{\bibfnamefont{M.}~\bibnamefont{Dal~Peraro}},
  \bibinfo{journal}{J. Chem. Theory Comput.} \textbf{\bibinfo{volume}{6}},
  \bibinfo{pages}{315} (\bibinfo{year}{2010}).

\bibitem[{\citenamefont{Enciso and Rey}(2010)}]{Enciso2010}
\bibinfo{author}{\bibfnamefont{M.}~\bibnamefont{Enciso}} \bibnamefont{and}
  \bibinfo{author}{\bibfnamefont{A.}~\bibnamefont{Rey}}, \bibinfo{journal}{J.
  Chem. Phys.} \textbf{\bibinfo{volume}{132}}, \bibinfo{pages}{235102}
  (\bibinfo{year}{2010}).

\bibitem[{\citenamefont{Enciso and Rey}(2012)}]{Enciso2012}
\bibinfo{author}{\bibfnamefont{M.}~\bibnamefont{Enciso}} \bibnamefont{and}
  \bibinfo{author}{\bibfnamefont{A.}~\bibnamefont{Rey}}, \bibinfo{journal}{J.
  Chem. Phys.} \textbf{\bibinfo{volume}{136}}, \bibinfo{pages}{215103}
  (\bibinfo{year}{2012}).

\bibitem[{\citenamefont{Spiga et~al.}(2013)\citenamefont{Spiga, Alemani,
  Degiacomi, Cascella, and Dal~Peraro}}]{Spiga2013}
\bibinfo{author}{\bibfnamefont{E.}~\bibnamefont{Spiga}},
  \bibinfo{author}{\bibfnamefont{D.}~\bibnamefont{Alemani}},
  \bibinfo{author}{\bibfnamefont{M.~T.} \bibnamefont{Degiacomi}},
  \bibinfo{author}{\bibfnamefont{M.}~\bibnamefont{Cascella}}, \bibnamefont{and}
  \bibinfo{author}{\bibfnamefont{M.}~\bibnamefont{Dal~Peraro}},
  \bibinfo{journal}{J. Chem. Theory Comput.} \textbf{\bibinfo{volume}{9}},
  \bibinfo{pages}{3515} (\bibinfo{year}{2013}).

\bibitem[{\citenamefont{Ouldridge et~al.}(2010)\citenamefont{Ouldridge, Louis,
  and Doye}}]{Ouldridge2010}
\bibinfo{author}{\bibfnamefont{T.~E.} \bibnamefont{Ouldridge}},
  \bibinfo{author}{\bibfnamefont{A.~A.} \bibnamefont{Louis}}, \bibnamefont{and}
  \bibinfo{author}{\bibfnamefont{J.~P.~K.} \bibnamefont{Doye}},
  \bibinfo{journal}{Phys. Rev. Lett.} \textbf{\bibinfo{volume}{104}},
  \bibinfo{pages}{178101} (\bibinfo{year}{2010}).

\bibitem[{\citenamefont{Morriss-Andrews
  et~al.}(2010)\citenamefont{Morriss-Andrews, Rottler, and
  Plotkin}}]{Morriss-Andrews2010}
\bibinfo{author}{\bibfnamefont{A.}~\bibnamefont{Morriss-Andrews}},
  \bibinfo{author}{\bibfnamefont{J.}~\bibnamefont{Rottler}}, \bibnamefont{and}
  \bibinfo{author}{\bibfnamefont{S.~S.} \bibnamefont{Plotkin}},
  \bibinfo{journal}{J. Chem. Phys.} \textbf{\bibinfo{volume}{132}},
  \bibinfo{pages}{035105} (\bibinfo{year}{2010}).

\bibitem[{\citenamefont{Linak et~al.}(2011)\citenamefont{Linak, Tourdot, and
  Dorfman}}]{Linak2011}
\bibinfo{author}{\bibfnamefont{M.~C.} \bibnamefont{Linak}},
  \bibinfo{author}{\bibfnamefont{R.}~\bibnamefont{Tourdot}}, \bibnamefont{and}
  \bibinfo{author}{\bibfnamefont{K.~D.} \bibnamefont{Dorfman}},
  \bibinfo{journal}{J. Chem. Phys.} \textbf{\bibinfo{volume}{135}},
  \bibinfo{pages}{205102} (\bibinfo{year}{2011}).

\bibitem[{\citenamefont{Sulc et~al.}(2012)\citenamefont{Sulc, Romano,
  Ouldridge, Rovigatti, Doye, and Louis}}]{Sulc2013}
\bibinfo{author}{\bibfnamefont{P.}~\bibnamefont{Sulc}},
  \bibinfo{author}{\bibfnamefont{F.}~\bibnamefont{Romano}},
  \bibinfo{author}{\bibfnamefont{T.~E.} \bibnamefont{Ouldridge}},
  \bibinfo{author}{\bibfnamefont{L.}~\bibnamefont{Rovigatti}},
  \bibinfo{author}{\bibfnamefont{J.~P.~K.} \bibnamefont{Doye}},
  \bibnamefont{and} \bibinfo{author}{\bibfnamefont{A.~A.} \bibnamefont{Louis}},
  \bibinfo{journal}{J. Chem. Phys.} \textbf{\bibinfo{volume}{137}},
  \bibinfo{pages}{135101} (\bibinfo{year}{2012}).

\bibitem[{\citenamefont{Orsi and Essex}(2011)}]{Orsi2011}
\bibinfo{author}{\bibfnamefont{M.}~\bibnamefont{Orsi}} \bibnamefont{and}
  \bibinfo{author}{\bibfnamefont{J.~W.} \bibnamefont{Essex}},
  \bibinfo{journal}{PLoS ONE} \textbf{\bibinfo{volume}{6}},
  \bibinfo{pages}{e28637} (\bibinfo{year}{2011}).

\bibitem[{\citenamefont{Ayton et~al.}(2007)\citenamefont{Ayton, Noid, and
  Voth}}]{Ayton2007}
\bibinfo{author}{\bibfnamefont{G.~S.} \bibnamefont{Ayton}},
  \bibinfo{author}{\bibfnamefont{W.~G.} \bibnamefont{Noid}}, \bibnamefont{and}
  \bibinfo{author}{\bibfnamefont{G.~A.} \bibnamefont{Voth}},
  \bibinfo{journal}{Curr. Opin. Struct. Biol.} \textbf{\bibinfo{volume}{17}},
  \bibinfo{pages}{192} (\bibinfo{year}{2007}).

\bibitem[{\citenamefont{Praprotnik et~al.}(2008)\citenamefont{Praprotnik,
  Delle~Site, and Kremer}}]{Praprotnik2008}
\bibinfo{author}{\bibfnamefont{M.}~\bibnamefont{Praprotnik}},
  \bibinfo{author}{\bibfnamefont{L.}~\bibnamefont{Delle~Site}},
  \bibnamefont{and} \bibinfo{author}{\bibfnamefont{K.}~\bibnamefont{Kremer}},
  \bibinfo{journal}{Annu. Rev. Phys. Chem.} \textbf{\bibinfo{volume}{59}},
  \bibinfo{pages}{545} (\bibinfo{year}{2008}).

\bibitem[{\citenamefont{Peter and Kremer}(2009)}]{Peter2009}
\bibinfo{author}{\bibfnamefont{C.}~\bibnamefont{Peter}} \bibnamefont{and}
  \bibinfo{author}{\bibfnamefont{K.}~\bibnamefont{Kremer}},
  \bibinfo{journal}{Soft Matter} \textbf{\bibinfo{volume}{5}},
  \bibinfo{pages}{4347} (\bibinfo{year}{2009}).

\bibitem[{\citenamefont{Meier et~al.}(2013)\citenamefont{Meier, Choutko,
  Dolenc, Eichenberger, Riniker, and van Gunsteren}}]{Meier2013}
\bibinfo{author}{\bibfnamefont{K.}~\bibnamefont{Meier}},
  \bibinfo{author}{\bibfnamefont{A.}~\bibnamefont{Choutko}},
  \bibinfo{author}{\bibfnamefont{J.}~\bibnamefont{Dolenc}},
  \bibinfo{author}{\bibfnamefont{A.~P.} \bibnamefont{Eichenberger}},
  \bibinfo{author}{\bibfnamefont{S.}~\bibnamefont{Riniker}}, \bibnamefont{and}
  \bibinfo{author}{\bibfnamefont{W.~F.} \bibnamefont{van Gunsteren}},
  \bibinfo{journal}{Angew. Chem. Int. Ed.} \textbf{\bibinfo{volume}{52}},
  \bibinfo{pages}{2820} (\bibinfo{year}{2013}).

\bibitem[{\citenamefont{Haxton}()}]{Haxtonproteininprep}
\bibinfo{author}{\bibfnamefont{T.~K.} \bibnamefont{Haxton}}, \bibinfo{note}{in
  preparation}.

\bibitem[{\citenamefont{Wolfenden et~al.}(1981)\citenamefont{Wolfenden,
  Andersson, Cullis, and Southgate}}]{Wolfenden1981}
\bibinfo{author}{\bibfnamefont{R.}~\bibnamefont{Wolfenden}},
  \bibinfo{author}{\bibfnamefont{L.}~\bibnamefont{Andersson}},
  \bibinfo{author}{\bibfnamefont{P.~M.} \bibnamefont{Cullis}},
  \bibnamefont{and} \bibinfo{author}{\bibfnamefont{C.~C.~B.}
  \bibnamefont{Southgate}}, \bibinfo{journal}{Biochemistry}
  \textbf{\bibinfo{volume}{20}}, \bibinfo{pages}{849} (\bibinfo{year}{1981}).

\bibitem[{\citenamefont{K\:{o}nig et~al.}(2013)\citenamefont{K\:{o}nig,
  Bruckner, and Boresch}}]{Konig2013}
\bibinfo{author}{\bibfnamefont{G.}~\bibnamefont{K\:{o}nig}},
  \bibinfo{author}{\bibfnamefont{S.}~\bibnamefont{Bruckner}}, \bibnamefont{and}
  \bibinfo{author}{\bibfnamefont{S.}~\bibnamefont{Boresch}},
  \bibinfo{journal}{Biophys. J.} \textbf{\bibinfo{volume}{104}},
  \bibinfo{pages}{453} (\bibinfo{year}{2013}).

\bibitem[{\citenamefont{Mannige et~al.}()\citenamefont{Mannige, Haxton, Proulx,
  Butterfoss, , Zuckermann, and Whitelam}}]{Mannigeinprep}
\bibinfo{author}{\bibfnamefont{R.~V.} \bibnamefont{Mannige}},
  \bibinfo{author}{\bibfnamefont{T.~K.} \bibnamefont{Haxton}},
  \bibinfo{author}{\bibfnamefont{C.}~\bibnamefont{Proulx}},
  \bibinfo{author}{\bibfnamefont{G.~L.} \bibnamefont{Butterfoss}}, ,
  \bibinfo{author}{\bibfnamefont{R.~N.} \bibnamefont{Zuckermann}},
  \bibnamefont{and} \bibinfo{author}{\bibfnamefont{S.}~\bibnamefont{Whitelam}},
  \bibinfo{note}{submitted}.

\bibitem[{\citenamefont{Murnen et~al.}(2012)\citenamefont{Murnen, Khokhlov,
  Khalatur, Segalman, and Zuckermann}}]{Murnen2012}
\bibinfo{author}{\bibfnamefont{H.~K.} \bibnamefont{Murnen}},
  \bibinfo{author}{\bibfnamefont{A.~R.} \bibnamefont{Khokhlov}},
  \bibinfo{author}{\bibfnamefont{P.~G.} \bibnamefont{Khalatur}},
  \bibinfo{author}{\bibfnamefont{R.~A.} \bibnamefont{Segalman}},
  \bibnamefont{and} \bibinfo{author}{\bibfnamefont{R.~N.}
  \bibnamefont{Zuckermann}}, \bibinfo{journal}{Macromolecules}
  \textbf{\bibinfo{volume}{45}}, \bibinfo{pages}{5229} (\bibinfo{year}{2012}).

\bibitem[{\citenamefont{Mylonas and Svergun}(2007)}]{Mylonas2007}
\bibinfo{author}{\bibfnamefont{E.}~\bibnamefont{Mylonas}} \bibnamefont{and}
  \bibinfo{author}{\bibfnamefont{D.~I.} \bibnamefont{Svergun}},
  \bibinfo{journal}{J. Appl. Cryst.} \textbf{\bibinfo{volume}{40}},
  \bibinfo{pages}{s245} (\bibinfo{year}{2007}).

\bibitem[{\citenamefont{Tsch\"op et~al.}(1998)\citenamefont{Tsch\"op, Kremer,
  Hahn, Batoulis, and B\"urger}}]{Tschop1998}
\bibinfo{author}{\bibfnamefont{W.}~\bibnamefont{Tsch\"op}},
  \bibinfo{author}{\bibfnamefont{K.}~\bibnamefont{Kremer}},
  \bibinfo{author}{\bibfnamefont{O.}~\bibnamefont{Hahn}},
  \bibinfo{author}{\bibfnamefont{J.}~\bibnamefont{Batoulis}}, \bibnamefont{and}
  \bibinfo{author}{\bibfnamefont{T.}~\bibnamefont{B\"urger}},
  \bibinfo{journal}{Acta Polymer.} \textbf{\bibinfo{volume}{49}},
  \bibinfo{pages}{75} (\bibinfo{year}{1998}).

\bibitem[{\citenamefont{Shih et~al.}(2007)\citenamefont{Shih, Freddolino,
  Sligar, and Schulten}}]{Shih2007}
\bibinfo{author}{\bibfnamefont{A.~Y.} \bibnamefont{Shih}},
  \bibinfo{author}{\bibfnamefont{P.~L.} \bibnamefont{Freddolino}},
  \bibinfo{author}{\bibfnamefont{S.~G.} \bibnamefont{Sligar}},
  \bibnamefont{and} \bibinfo{author}{\bibfnamefont{K.}~\bibnamefont{Schulten}},
  \bibinfo{journal}{Nano Lett.} \textbf{\bibinfo{volume}{7}},
  \bibinfo{pages}{1692} (\bibinfo{year}{2007}).

\bibitem[{\citenamefont{Perlmutter and Sachs}(2009)}]{Perlmutter2009}
\bibinfo{author}{\bibfnamefont{J.~D.} \bibnamefont{Perlmutter}}
  \bibnamefont{and} \bibinfo{author}{\bibfnamefont{J.~N.} \bibnamefont{Sachs}},
  \bibinfo{journal}{Biochim. Biophys. Acta} \textbf{\bibinfo{volume}{1788}},
  \bibinfo{pages}{2284} (\bibinfo{year}{2009}).

\bibitem[{\citenamefont{Rosales et~al.}(2012)\citenamefont{Rosales, Murnen,
  Kline, Zuckermann, and Segalman}}]{Rosales2012}
\bibinfo{author}{\bibfnamefont{A.~R.} \bibnamefont{Rosales}},
  \bibinfo{author}{\bibfnamefont{H.~K.} \bibnamefont{Murnen}},
  \bibinfo{author}{\bibfnamefont{S.~R.} \bibnamefont{Kline}},
  \bibinfo{author}{\bibfnamefont{R.~N.} \bibnamefont{Zuckermann}},
  \bibnamefont{and} \bibinfo{author}{\bibfnamefont{R.~A.}
  \bibnamefont{Segalman}}, \bibinfo{journal}{Soft Matter}
  \textbf{\bibinfo{volume}{8}}, \bibinfo{pages}{3673} (\bibinfo{year}{2012}).

\bibitem[{\citenamefont{Murnen et~al.}(2013)\citenamefont{Murnen, Rosales,
  Dobrynin, Zuckermann, and Segalman}}]{Murnen2013}
\bibinfo{author}{\bibfnamefont{H.~K.} \bibnamefont{Murnen}},
  \bibinfo{author}{\bibfnamefont{A.~M.} \bibnamefont{Rosales}},
  \bibinfo{author}{\bibfnamefont{A.~V.} \bibnamefont{Dobrynin}},
  \bibinfo{author}{\bibfnamefont{R.~N.} \bibnamefont{Zuckermann}},
  \bibnamefont{and} \bibinfo{author}{\bibfnamefont{R.~A.}
  \bibnamefont{Segalman}}, \bibinfo{journal}{Soft Matter}
  \textbf{\bibinfo{volume}{9}}, \bibinfo{pages}{90} (\bibinfo{year}{2013}).

\bibitem[{\citenamefont{Lozinsky}(2006)}]{Lozinsky2006}
\bibinfo{author}{\bibfnamefont{V.~I.} \bibnamefont{Lozinsky}},
  \bibinfo{journal}{Adv. Polym. Sci.} \textbf{\bibinfo{volume}{196}},
  \bibinfo{pages}{87} (\bibinfo{year}{2006}).

\bibitem[{\citenamefont{Altintas and Barner-Kowollik}(2012)}]{Altintas2012}
\bibinfo{author}{\bibfnamefont{O.}~\bibnamefont{Altintas}} \bibnamefont{and}
  \bibinfo{author}{\bibfnamefont{C.}~\bibnamefont{Barner-Kowollik}},
  \bibinfo{journal}{Macromol. Rapid Comm.} \textbf{\bibinfo{volume}{33}},
  \bibinfo{pages}{958} (\bibinfo{year}{2012}).

\bibitem[{\citenamefont{Akagi et~al.}(2012)\citenamefont{Akagi, Piyapakorn, and
  Akashi}}]{Akagi2012}
\bibinfo{author}{\bibfnamefont{T.}~\bibnamefont{Akagi}},
  \bibinfo{author}{\bibfnamefont{P.}~\bibnamefont{Piyapakorn}},
  \bibnamefont{and} \bibinfo{author}{\bibfnamefont{M.}~\bibnamefont{Akashi}},
  \bibinfo{journal}{Langmuir} \textbf{\bibinfo{volume}{28}},
  \bibinfo{pages}{5249} (\bibinfo{year}{2012}).

\bibitem[{\citenamefont{Moreno et~al.}(2013)\citenamefont{Moreno, Lo~Verso,
  Sanchez-Sanchez, Arbe, Colmenero, and Pomposo}}]{Morena2013}
\bibinfo{author}{\bibfnamefont{A.~J.} \bibnamefont{Moreno}},
  \bibinfo{author}{\bibfnamefont{F.}~\bibnamefont{Lo~Verso}},
  \bibinfo{author}{\bibfnamefont{A.}~\bibnamefont{Sanchez-Sanchez}},
  \bibinfo{author}{\bibfnamefont{A.}~\bibnamefont{Arbe}},
  \bibinfo{author}{\bibfnamefont{J.}~\bibnamefont{Colmenero}},
  \bibnamefont{and} \bibinfo{author}{\bibfnamefont{J.~A.}
  \bibnamefont{Pomposo}}, \bibinfo{journal}{Macromolecules}
  \textbf{\bibinfo{volume}{46}}, \bibinfo{pages}{9748} (\bibinfo{year}{2013}).

\bibitem[{\citenamefont{Lo~Verso et~al.}(2014)\citenamefont{Lo~Verso, Pomposo,
  Colmenero, and Moreno}}]{LoVerso2014}
\bibinfo{author}{\bibfnamefont{F.}~\bibnamefont{Lo~Verso}},
  \bibinfo{author}{\bibfnamefont{J.~A.} \bibnamefont{Pomposo}},
  \bibinfo{author}{\bibfnamefont{J.}~\bibnamefont{Colmenero}},
  \bibnamefont{and} \bibinfo{author}{\bibfnamefont{A.~J.}
  \bibnamefont{Moreno}}, \bibinfo{journal}{Soft Matter}
  \textbf{\bibinfo{volume}{10}}, \bibinfo{pages}{4813} (\bibinfo{year}{2014}).

\bibitem[{\citenamefont{Terashima et~al.}(2014)\citenamefont{Terashima, Sugita,
  Fukae, and Sawamotot}}]{Terashima2014}
\bibinfo{author}{\bibfnamefont{T.}~\bibnamefont{Terashima}},
  \bibinfo{author}{\bibfnamefont{T.}~\bibnamefont{Sugita}},
  \bibinfo{author}{\bibfnamefont{K.}~\bibnamefont{Fukae}}, \bibnamefont{and}
  \bibinfo{author}{\bibfnamefont{M.}~\bibnamefont{Sawamotot}},
  \bibinfo{journal}{Macromolecules} \textbf{\bibinfo{volume}{47}},
  \bibinfo{pages}{589} (\bibinfo{year}{2014}).

\bibitem[{\citenamefont{Neri et~al.}(2005)\citenamefont{Neri, Anselmi,
  Cascella, Maritan, and Carloni}}]{Neri2005}
\bibinfo{author}{\bibfnamefont{M.}~\bibnamefont{Neri}},
  \bibinfo{author}{\bibfnamefont{C.}~\bibnamefont{Anselmi}},
  \bibinfo{author}{\bibfnamefont{M.}~\bibnamefont{Cascella}},
  \bibinfo{author}{\bibfnamefont{A.}~\bibnamefont{Maritan}}, \bibnamefont{and}
  \bibinfo{author}{\bibfnamefont{P.}~\bibnamefont{Carloni}},
  \bibinfo{journal}{Phys. Rev. Lett.} \textbf{\bibinfo{volume}{95}},
  \bibinfo{pages}{218102} (\bibinfo{year}{2005}).

\bibitem[{\citenamefont{Shi et~al.}(2006)\citenamefont{Shi, Izvekov, and
  Voth}}]{Shi2006}
\bibinfo{author}{\bibfnamefont{Q.}~\bibnamefont{Shi}},
  \bibinfo{author}{\bibfnamefont{S.}~\bibnamefont{Izvekov}}, \bibnamefont{and}
  \bibinfo{author}{\bibfnamefont{G.~A.} \bibnamefont{Voth}},
  \bibinfo{journal}{J. Phys. Chem. B} \textbf{\bibinfo{volume}{110}},
  \bibinfo{pages}{15045} (\bibinfo{year}{2006}).

\bibitem[{\citenamefont{Machado et~al.}(2011)\citenamefont{Machado, Dans, and
  Pantano}}]{Machado2011}
\bibinfo{author}{\bibfnamefont{M.~R.} \bibnamefont{Machado}},
  \bibinfo{author}{\bibfnamefont{P.~D.} \bibnamefont{Dans}}, \bibnamefont{and}
  \bibinfo{author}{\bibfnamefont{S.}~\bibnamefont{Pantano}},
  \bibinfo{journal}{Phys. Chem. Chem. Phys.} \textbf{\bibinfo{volume}{13}},
  \bibinfo{pages}{18134} (\bibinfo{year}{2011}).

\bibitem[{\citenamefont{Mamonov et~al.}(2012)\citenamefont{Mamonov, Lettieri,
  Ding, Sarver, Palli, Cunningham, Saxena, and Zuckerman}}]{Mamonov2012}
\bibinfo{author}{\bibfnamefont{A.~B.} \bibnamefont{Mamonov}},
  \bibinfo{author}{\bibfnamefont{S.}~\bibnamefont{Lettieri}},
  \bibinfo{author}{\bibfnamefont{Y.}~\bibnamefont{Ding}},
  \bibinfo{author}{\bibfnamefont{J.~L.} \bibnamefont{Sarver}},
  \bibinfo{author}{\bibfnamefont{R.}~\bibnamefont{Palli}},
  \bibinfo{author}{\bibfnamefont{T.~F.} \bibnamefont{Cunningham}},
  \bibinfo{author}{\bibfnamefont{S.}~\bibnamefont{Saxena}}, \bibnamefont{and}
  \bibinfo{author}{\bibfnamefont{D.~M.} \bibnamefont{Zuckerman}},
  \bibinfo{journal}{J. Chem. Theory Comput.} \textbf{\bibinfo{volume}{8}},
  \bibinfo{pages}{2921} (\bibinfo{year}{2012}).

\bibitem[{\citenamefont{di~Pasquale et~al.}(2012)\citenamefont{di~Pasquale,
  Marchisio, and Carbone}}]{diPasquale2012}
\bibinfo{author}{\bibfnamefont{N.}~\bibnamefont{di~Pasquale}},
  \bibinfo{author}{\bibfnamefont{D.}~\bibnamefont{Marchisio}},
  \bibnamefont{and} \bibinfo{author}{\bibfnamefont{P.}~\bibnamefont{Carbone}},
  \bibinfo{journal}{J. Chem. Phys.} \textbf{\bibinfo{volume}{137}},
  \bibinfo{pages}{164111} (\bibinfo{year}{2012}).

\bibitem[{\citenamefont{Lyman et~al.}(2006)\citenamefont{Lyman, Ytreberg, and
  Zuckerman}}]{Lyman2006}
\bibinfo{author}{\bibfnamefont{E.}~\bibnamefont{Lyman}},
  \bibinfo{author}{\bibfnamefont{F.~M.} \bibnamefont{Ytreberg}},
  \bibnamefont{and} \bibinfo{author}{\bibfnamefont{D.~M.}
  \bibnamefont{Zuckerman}}, \bibinfo{journal}{Phys. Rev. Lett.}
  \textbf{\bibinfo{volume}{96}}, \bibinfo{pages}{028105}
  (\bibinfo{year}{2006}).

\bibitem[{\citenamefont{Lyman and Zuckerman}(2006)}]{Lyman2006b}
\bibinfo{author}{\bibfnamefont{E.}~\bibnamefont{Lyman}} \bibnamefont{and}
  \bibinfo{author}{\bibfnamefont{D.~M.} \bibnamefont{Zuckerman}},
  \bibinfo{journal}{J. Chem. Theory Comput.} \textbf{\bibinfo{volume}{2}},
  \bibinfo{pages}{656} (\bibinfo{year}{2006}).

\bibitem[{\citenamefont{Christen and van Gunsteren}(2006)}]{Christen2006}
\bibinfo{author}{\bibfnamefont{M.}~\bibnamefont{Christen}} \bibnamefont{and}
  \bibinfo{author}{\bibfnamefont{W.~F.} \bibnamefont{van Gunsteren}},
  \bibinfo{journal}{J. Chem. Phys.} \textbf{\bibinfo{volume}{124}},
  \bibinfo{pages}{154106} (\bibinfo{year}{2006}).

\bibitem[{\citenamefont{Moritsugu et~al.}(2010)\citenamefont{Moritsugu, Terada,
  and Kidera}}]{Moritsugu2010}
\bibinfo{author}{\bibfnamefont{K.}~\bibnamefont{Moritsugu}},
  \bibinfo{author}{\bibfnamefont{T.}~\bibnamefont{Terada}}, \bibnamefont{and}
  \bibinfo{author}{\bibfnamefont{A.}~\bibnamefont{Kidera}},
  \bibinfo{journal}{J. Chem. Phys.} \textbf{\bibinfo{volume}{133}},
  \bibinfo{pages}{244105} (\bibinfo{year}{2010}).

\bibitem[{\citenamefont{Thogersen et~al.}(2008)\citenamefont{Thogersen,
  Schiott, Vosegaard, Nielsen, and Tajkhorshid}}]{Thogersen2008}
\bibinfo{author}{\bibfnamefont{L.}~\bibnamefont{Thogersen}},
  \bibinfo{author}{\bibfnamefont{B.}~\bibnamefont{Schiott}},
  \bibinfo{author}{\bibfnamefont{T.}~\bibnamefont{Vosegaard}},
  \bibinfo{author}{\bibfnamefont{N.~C.} \bibnamefont{Nielsen}},
  \bibnamefont{and}
  \bibinfo{author}{\bibfnamefont{E.}~\bibnamefont{Tajkhorshid}},
  \bibinfo{journal}{Biophys. J.} \textbf{\bibinfo{volume}{95}},
  \bibinfo{pages}{4337} (\bibinfo{year}{2008}).

\bibitem[{\citenamefont{Stansfeld and Sansom}(2011)}]{Stansfeld2011}
\bibinfo{author}{\bibfnamefont{P.~J.} \bibnamefont{Stansfeld}}
  \bibnamefont{and} \bibinfo{author}{\bibfnamefont{M.~S.~P.}
  \bibnamefont{Sansom}}, \bibinfo{journal}{J. Chem. Theory Comput.}
  \textbf{\bibinfo{volume}{7}}, \bibinfo{pages}{1157} (\bibinfo{year}{2011}).

\bibitem[{\citenamefont{Permutter et~al.}(2011)\citenamefont{Permutter,
  Drasler, Xie, Gao, Popot, and Sachs}}]{Perlmutter2011}
\bibinfo{author}{\bibfnamefont{J.~D.} \bibnamefont{Permutter}},
  \bibinfo{author}{\bibfnamefont{W.~J.} \bibnamefont{Drasler}},
  \bibinfo{author}{\bibfnamefont{W.}~\bibnamefont{Xie}},
  \bibinfo{author}{\bibfnamefont{L.}~\bibnamefont{Gao}},
  \bibinfo{author}{\bibfnamefont{J.-L.} \bibnamefont{Popot}}, \bibnamefont{and}
  \bibinfo{author}{\bibfnamefont{J.~N.} \bibnamefont{Sachs}},
  \bibinfo{journal}{Langmuir} \textbf{\bibinfo{volume}{27}},
  \bibinfo{pages}{10523} (\bibinfo{year}{2011}).

\bibitem[{\citenamefont{Mackerell et~al.}(2004)\citenamefont{Mackerell, Feig,
  and Brooks}}]{CHARMM}
\bibinfo{author}{\bibfnamefont{A.~D.} \bibnamefont{Mackerell}},
  \bibinfo{author}{\bibfnamefont{M.}~\bibnamefont{Feig}}, \bibnamefont{and}
  \bibinfo{author}{\bibfnamefont{C.~L.} \bibnamefont{Brooks}},
  \bibinfo{journal}{J. Comput. Chem.} \textbf{\bibinfo{volume}{25}},
  \bibinfo{pages}{1400} (\bibinfo{year}{2004}).

\bibitem[{\citenamefont{Mirijanian et~al.}(2014)\citenamefont{Mirijanian,
  Mannige, Zuckermann, and Whitelam}}]{Mirijanian2013}
\bibinfo{author}{\bibfnamefont{D.~T.} \bibnamefont{Mirijanian}},
  \bibinfo{author}{\bibfnamefont{R.~V.} \bibnamefont{Mannige}},
  \bibinfo{author}{\bibfnamefont{R.~N.} \bibnamefont{Zuckermann}},
  \bibnamefont{and} \bibinfo{author}{\bibfnamefont{S.}~\bibnamefont{Whitelam}},
  \bibinfo{journal}{J. Comput. Chem.} \textbf{\bibinfo{volume}{35}},
  \bibinfo{pages}{360} (\bibinfo{year}{2014}).

\bibitem[{\citenamefont{Feig et~al.}(2004)\citenamefont{Feig, Karanicolas, and
  Brooks}}]{Feig2004}
\bibinfo{author}{\bibfnamefont{M.}~\bibnamefont{Feig}},
  \bibinfo{author}{\bibfnamefont{J.}~\bibnamefont{Karanicolas}},
  \bibnamefont{and} \bibinfo{author}{\bibfnamefont{C.~L.}
  \bibnamefont{Brooks}}, \bibinfo{journal}{J. Mol. Graph. Model.}
  \textbf{\bibinfo{volume}{22}}, \bibinfo{pages}{377} (\bibinfo{year}{2004}).

\bibitem[{\citenamefont{Berman et~al.}(2000)\citenamefont{Berman, Westbrook,
  Feng, Gilliland, Bhat, Weissig, Shindyalov, and Bourne}}]{PDB}
\bibinfo{author}{\bibfnamefont{H.~M.} \bibnamefont{Berman}},
  \bibinfo{author}{\bibfnamefont{J.}~\bibnamefont{Westbrook}},
  \bibinfo{author}{\bibfnamefont{Z.}~\bibnamefont{Feng}},
  \bibinfo{author}{\bibfnamefont{G.}~\bibnamefont{Gilliland}},
  \bibinfo{author}{\bibfnamefont{T.~N.} \bibnamefont{Bhat}},
  \bibinfo{author}{\bibfnamefont{H.}~\bibnamefont{Weissig}},
  \bibinfo{author}{\bibfnamefont{I.~N.} \bibnamefont{Shindyalov}},
  \bibnamefont{and} \bibinfo{author}{\bibfnamefont{P.~E.}
  \bibnamefont{Bourne}}, \bibinfo{journal}{Nucleic Acids Res.}
  \textbf{\bibinfo{volume}{28}}, \bibinfo{pages}{235} (\bibinfo{year}{2000}).

\bibitem[{\citenamefont{Zuckermann and Kodadek}(2009)}]{Zuckermann2009}
\bibinfo{author}{\bibfnamefont{R.~N.} \bibnamefont{Zuckermann}}
  \bibnamefont{and} \bibinfo{author}{\bibfnamefont{T.}~\bibnamefont{Kodadek}},
  \bibinfo{journal}{Curr. Opin. Mol. Ther.} \textbf{\bibinfo{volume}{11}},
  \bibinfo{pages}{299} (\bibinfo{year}{2009}).

\bibitem[{\citenamefont{Gao et~al.}(2010)\citenamefont{Gao, Yam, Wang,
  Magdangal, Salisbury, Peretz, Zuckermann, Connolly, Hansson, and
  Minthon}}]{Gao2010}
\bibinfo{author}{\bibfnamefont{C.~M.} \bibnamefont{Gao}},
  \bibinfo{author}{\bibfnamefont{A.~Y.} \bibnamefont{Yam}},
  \bibinfo{author}{\bibfnamefont{X.}~\bibnamefont{Wang}},
  \bibinfo{author}{\bibfnamefont{E.}~\bibnamefont{Magdangal}},
  \bibinfo{author}{\bibfnamefont{C.}~\bibnamefont{Salisbury}},
  \bibinfo{author}{\bibfnamefont{D.}~\bibnamefont{Peretz}},
  \bibinfo{author}{\bibfnamefont{R.~N.} \bibnamefont{Zuckermann}},
  \bibinfo{author}{\bibfnamefont{M.~D.} \bibnamefont{Connolly}},
  \bibinfo{author}{\bibfnamefont{O.}~\bibnamefont{Hansson}}, \bibnamefont{and}
  \bibinfo{author}{\bibfnamefont{L.}~\bibnamefont{Minthon}},
  \bibinfo{journal}{PLOS ONE} p. \bibinfo{pages}{e15725}
  (\bibinfo{year}{2010}).

\bibitem[{\citenamefont{Yam et~al.}(2011)\citenamefont{Yam, Wang, Gao,
  Connolly, Zuckermann, Bleu, Hall, Fedynyshyn, Allauzen, and
  Peretz}}]{Yam2011}
\bibinfo{author}{\bibfnamefont{A.~Y.} \bibnamefont{Yam}},
  \bibinfo{author}{\bibfnamefont{X.}~\bibnamefont{Wang}},
  \bibinfo{author}{\bibfnamefont{C.~M.} \bibnamefont{Gao}},
  \bibinfo{author}{\bibfnamefont{M.~D.} \bibnamefont{Connolly}},
  \bibinfo{author}{\bibfnamefont{R.~N.} \bibnamefont{Zuckermann}},
  \bibinfo{author}{\bibfnamefont{T.}~\bibnamefont{Bleu}},
  \bibinfo{author}{\bibfnamefont{J.}~\bibnamefont{Hall}},
  \bibinfo{author}{\bibfnamefont{J.~P.} \bibnamefont{Fedynyshyn}},
  \bibinfo{author}{\bibfnamefont{S.}~\bibnamefont{Allauzen}}, \bibnamefont{and}
  \bibinfo{author}{\bibfnamefont{D.}~\bibnamefont{Peretz}},
  \bibinfo{journal}{Biochemistry} \textbf{\bibinfo{volume}{50}},
  \bibinfo{pages}{4322} (\bibinfo{year}{2011}).

\bibitem[{\citenamefont{Reddy et~al.}(2011)\citenamefont{Reddy, Wilson, Wilson,
  Connell, Gocke, Hynan, German, and Kodadek}}]{Reddy2011}
\bibinfo{author}{\bibfnamefont{M.~M.} \bibnamefont{Reddy}},
  \bibinfo{author}{\bibfnamefont{R.}~\bibnamefont{Wilson}},
  \bibinfo{author}{\bibfnamefont{J.}~\bibnamefont{Wilson}},
  \bibinfo{author}{\bibfnamefont{S.}~\bibnamefont{Connell}},
  \bibinfo{author}{\bibfnamefont{A.}~\bibnamefont{Gocke}},
  \bibinfo{author}{\bibfnamefont{L.}~\bibnamefont{Hynan}},
  \bibinfo{author}{\bibfnamefont{D.}~\bibnamefont{German}}, \bibnamefont{and}
  \bibinfo{author}{\bibfnamefont{T.}~\bibnamefont{Kodadek}},
  \bibinfo{journal}{Cell} \textbf{\bibinfo{volume}{144}}, \bibinfo{pages}{132}
  (\bibinfo{year}{2011}).

\bibitem[{\citenamefont{Utku et~al.}(2006)\citenamefont{Utku, Dehan, Oeurfelli,
  Piano, Zuckermann, Pagano, and Kirshenbaum}}]{Utku2006}
\bibinfo{author}{\bibfnamefont{Y.}~\bibnamefont{Utku}},
  \bibinfo{author}{\bibfnamefont{E.}~\bibnamefont{Dehan}},
  \bibinfo{author}{\bibfnamefont{O.}~\bibnamefont{Oeurfelli}},
  \bibinfo{author}{\bibfnamefont{F.}~\bibnamefont{Piano}},
  \bibinfo{author}{\bibfnamefont{R.~N.} \bibnamefont{Zuckermann}},
  \bibinfo{author}{\bibfnamefont{M.}~\bibnamefont{Pagano}}, \bibnamefont{and}
  \bibinfo{author}{\bibfnamefont{K.~A.} \bibnamefont{Kirshenbaum}},
  \bibinfo{journal}{Mol. BioSyst.} \textbf{\bibinfo{volume}{2}},
  \bibinfo{pages}{312} (\bibinfo{year}{2006}).

\bibitem[{\citenamefont{Lobo et~al.}(2003)\citenamefont{Lobo, Vetro, Suich,
  Zuckermann, and Middaugh}}]{Lobo2003}
\bibinfo{author}{\bibfnamefont{B.~A.} \bibnamefont{Lobo}},
  \bibinfo{author}{\bibfnamefont{J.~A.} \bibnamefont{Vetro}},
  \bibinfo{author}{\bibfnamefont{D.~M.} \bibnamefont{Suich}},
  \bibinfo{author}{\bibfnamefont{R.~N.} \bibnamefont{Zuckermann}},
  \bibnamefont{and} \bibinfo{author}{\bibfnamefont{C.~R.}
  \bibnamefont{Middaugh}}, \bibinfo{journal}{J. Pharm. Sci.}
  \textbf{\bibinfo{volume}{92}}, \bibinfo{pages}{1905} (\bibinfo{year}{2003}).

\bibitem[{\citenamefont{Chongsiriwatana
  et~al.}(2008)\citenamefont{Chongsiriwatana, Patch, Czyzewksi, Dohm, Ivankin,
  Gidalevitz, Zuckermann, and Barron}}]{Chongsiriwatana2008}
\bibinfo{author}{\bibfnamefont{N.~P.} \bibnamefont{Chongsiriwatana}},
  \bibinfo{author}{\bibfnamefont{J.~A.} \bibnamefont{Patch}},
  \bibinfo{author}{\bibfnamefont{A.~M.} \bibnamefont{Czyzewksi}},
  \bibinfo{author}{\bibfnamefont{M.~T.} \bibnamefont{Dohm}},
  \bibinfo{author}{\bibfnamefont{A.}~\bibnamefont{Ivankin}},
  \bibinfo{author}{\bibfnamefont{D.}~\bibnamefont{Gidalevitz}},
  \bibinfo{author}{\bibfnamefont{R.~N.} \bibnamefont{Zuckermann}},
  \bibnamefont{and} \bibinfo{author}{\bibfnamefont{A.~E.}
  \bibnamefont{Barron}}, \bibinfo{journal}{Proc. Natl. Acad. Sci. U.S.A.}
  \textbf{\bibinfo{volume}{105}}, \bibinfo{pages}{2794} (\bibinfo{year}{2008}).

\end{thebibliography}

\end{document}